\documentclass[graybox]{svmult}
\usepackage[utf8]{inputenc}
\usepackage[T1]{fontenc}

\usepackage{soul,xcolor}

\usepackage{type1cm}        % activate if the above 3 fonts are
\usepackage{makeidx}         % allows index generation
\usepackage{graphicx}        % standard LaTeX graphics tool
\usepackage{multicol}        % used for the two-column index
\usepackage[bottom]{footmisc}% places footnotes at page bottom

\usepackage{newtxtext}       % 
\usepackage{newtxmath}       % selects Times Roman as basic font
\UseRawInputEncoding
\usepackage{physics}
\usepackage{amsmath}
\usepackage{comment}
\usepackage{esvect}

\usepackage{hyperref}
\hypersetup{
    colorlinks=true,
    linkcolor=blue,
    filecolor=magenta,      
    urlcolor=blue,
    citecolor=red,
    pdftitle={Entanglement review},
    }
\makeindex             % used for the subject index
\usepackage{mathtools}

\begin{document}

\setstcolor{blue}

\title*{Introduction to quantum entanglement in many-body systems}

\author{Anubhav Kumar Srivastava, Guillem M\"uller-Rigat, Maciej Lewenstein and Grzegorz Rajchel-Mieldzio\'c}

\institute{Anubhav Kumar Srivastava$^*$  \at ICFO - Institut de Ciencies Fotoniques, The Barcelona Institute of Science and Technology, 08860 Castelldefels (Barcelona), Spain, \email{anubhav.srivastava@icfo.eu}
\and Guillem M\"uller-Rigat$^*$ \at ICFO - Institut de Ciencies Fotoniques, The Barcelona Institute of Science and Technology, 08860 Castelldefels (Barcelona), Spain, \email{guillem.muller@icfo.eu}
\and Maciej Lewenstein \at ICFO - Institut de Ciencies Fotoniques, The Barcelona Institute of Science and Technology, 08860 Castelldefels (Barcelona), Spain, \at ICREA, Pg. Llu\'{\i}s Companys 23, 08010 Barcelona, Spain, \email{maciej.lewenstein@icfo.eu} 
\and Grzegorz Rajchel-Mieldzio\'c \at ICFO - Institut de Ciencies Fotoniques, The Barcelona Institute of Science and Technology, 08860 Castelldefels (Barcelona), Spain, 
\at NASK National Research Institute, ul. Kolska 12, 01-045 Warszawa, Poland, \email{grzegorzrajchel@gmail.com} \\
$^*$ equal contribution}

\maketitle

\abstract{The quantum mechanics formalism introduced new revolutionary concepts challenging our everyday perceptions. Arguably, quantum entanglement, which explains correlations that cannot be reproduced classically, is the most notable of them. Besides its fundamental aspect, entanglement is also a resource, fueling emergent technologies such as quantum simulators and computers. The purpose of this \textit{chapter} is to give a pedagogical introduction to the topic with a special emphasis on the multipartite scenario, i.e., entanglement distributed among many degrees of freedom. Due to the combinatorial complexity of this setting, particles can interact and become entangled in a plethora of ways, which we characterize here.
We start by providing the necessary mathematical tools and elementary concepts from entanglement theory. A part of this \textit{chapter} will be devoted to classifying and ordering entangled states. Then, we focus on various entanglement structures useful in condensed-matter theory such as tensor-network states or symmetric states useful for quantum-enhanced sensing. Finally, we discuss state-of-the-art methods to detect and certify such correlations in experiments,  with some relevant illustrative examples.}

\tableofcontents

\newpage
\section{Introduction}

While many entanglement reviews have already been written, the majority of them were written a decade ago (cf. \cite{Amico2008, Horodecki_2009, Laflorencie2016, bengtsson2016brief, Walter_2016}). To cover the recent discoveries, we provide a review with an easy and didactic introduction to the topic, and at the same time, we discuss some of the newest results in the field.
The present text is based on a set of lectures by Prof. Lewenstein, and earlier reviews of the subject by some of us~\cite{Tura2016,Das2016,Lewenstein2012}.

The main focus of this review is on multipartite entanglement, with experimentally viable setups, but we start with the discussion of the simplest and the most elementary bipartite case in Section~\ref{sec:2}. 
We introduce here the notion of bipartite entanglement and some conceptual topics and discuss its quantification and applications.

In Section~\ref{sec:3}, we focus on multipartite entanglement and the challenges associated with it. Here, after introducing the basics, we discuss the notion of partial separability with respect to a specific example of absolutely maximally entangled (AME) states. Furthermore, we consider the resource theory of entanglement to account for the permissible operations and the classification of quantum states in various entanglement classes based on these operations. We study the example of the simplest case of a multipartite system, i.e., a system of 3-qubits to study the entanglement classes and the invariants parameterizing such classes.

Section~\ref{sec:4} concerns the use and detection of many-body quantum entanglement. Here, we define the entanglement depth and talk about the useful and useless entanglement. Tensor network states are presented here, and we discuss entanglement area laws. We also talk in this section about scalable entanglement certification, using entanglement witnesses. Finally, we introduce Bell correlations and discuss their implications.

We conclude in Section~\ref{sec:5}, by listing some open problems that are discussed within the scope of this chapter, particularly in the context of entanglement theory.

This chapter is based on the lectures Maciej Lewenstein delivered during the
XXIII Training Courses in the Physics of Strongly Correlated Systems, conducted in October 2021, and organized by the International Institute for Advanced Scientific Studies ``E.R. Caianiello'' (IIASS), in collaboration with the Department of Physics, University of Salerno. Although the focus of the course was ``Trends and Platforms for Quantum Technologies'', Lewenstein's lectures focused explicitly on the introduction to entanglement in general, and in many-body systems in particular.

\section{Mathematical foundations and bipartite systems}\label{sec:2}

The microscopic world is governed by linear laws, such as the Schr\"odinger equation, in opposition to what we can perceive with everyday experience. 
Thus, an appropriate description is given of modeling quantum objects as states from a linear space equipped with a scalar product for distinguishability -- Hilbert space. 

\begin{definition}
    Hilbert space is a complex linear space $\mathcal{H}$, with the inner product\footnote{For mathematical precision, we add that $\mathcal{H}$ is complete in the sense of metric induced by the inner product.} written as $\braket{\cdot}$. 
\end{definition}

\noindent \textbf{Quantum states}.-- Any state of the system must necessarily be physical, i.e., upon measurement in any basis the probabilities must sum up to one. 
For \emph{pure} states, this condition can be stated as follows:

\begin{definition}
    A pure state of the quantum system $\ket{\psi}\in \mathcal{H}$ from a Hilbert space $\mathcal{H}$ is a normalized vector $\braket{\psi}=1$.
\end{definition} 
Description using vectors from a Hilbert space is useful because it carries a lot of physical meaning. 
For example, two states $\ket{\psi}$ and $\ket{\phi}$ can be perfectly distinguished if and only if they are orthogonal in the vectorial sense~\cite{Helstrom_1969}, $\braket{\psi}{\phi} = 0$. 
Not all vectorial properties are physical though. 
Two orthogonal states will remain orthogonal after multiplication by a constant phase.
As a more general rule, multiplying by a global phase $e^{i\varphi}$ does not change a state and is not detectable, $e^{i\varphi}\ket{\psi} \simeq \ket{\psi}$. \\

\noindent\textbf{Measuring quantum states}.-- If we want to characterize a state, it is necessary to measure it. There are two basic types of measurements in quantum information: projective and positive operator-valued measure (POVM). The first one can be treated as a projection into one of the orthonormal vectors and is generally more useful from an experimentalist's perspective. Since usually, projective measurements are sufficient in a laboratory setting (such as a quantum circuit), for the rest of this contribution we shall discuss only these measurements.\footnote{However, from the foundational point of view, both projective measurements and POVMs are equivalent~\cite{Nielsen2012}.} 

\begin{definition}
   A projective measurement $\Pi$ is a set of projectors $\{\hat{\pi}_a \}$, pair-wise orthogonal $\hat{\mathrm{\pi}}_a\hat{\mathrm{\pi}}_b = \delta_{ab}\hat{\pi}_b $ and complete,\footnote{We denote $\mathbb{I}$ the identity operator acting in $\mathcal{H}$.} $\sum_a\hat{\pi}_a = \mathbb{I}$. 
\end{definition} 

\noindent\textit{Quantum phenomena do not occur in a Hilbert space, they occur in a laboratory} \cite{Peres2002}. Therefore, we need a law to map quantum states to the probabilities they predict.

\begin{overview}{Born's rule}
 The probability of obtaining outcome $a$ in measuring $\Pi = \{\hat{\pi}_a \}$ on the state $\ket{\psi}$ is given by~\cite{Born_1926}: 
\begin{equation}
    p_a(\psi) = \bra{\psi}{\hat{\pi}}_a\ket{\psi} \ .
\end{equation}
From it, one can deduce various statistics. The most important of them are: 

\begin{itemize}
    \item \textit{Expectation} or \emph{mean value} of its associated observable $ \hat{O} = \sum_{a}a\hat{\pi}_a$, 
    \begin{equation}
        \langle\hat{O} \rangle = \bra{\psi}\hat{O}\ket{\psi} \ . 
    \end{equation}

    \item \textit{Variance} as a measure of the fluctuations with respect to the mean, 
    \begin{equation}
        (\Delta\hat{O})^2  = \langle (\hat{O} - \langle\hat{O} \rangle)^2 \rangle = \langle\hat{O}^2 \rangle - \langle \hat{O}\rangle^2 \ .
    \end{equation}
\end{itemize}
\end{overview}
\noindent Contrary to our everyday intuition, not all observables are jointly measurable in the same quantum system. If two observables $\hat{A}, \hat{B}$ do not commute, $[\hat{A},\hat{B}] := \hat{A}\hat{B}- \hat{B}\hat{A}\neq 0$, they are generally not well-defined simultaneously, i.e., they do not have a joint deterministic outcome. This unintuitive fact is enclosed in Heisenberg's uncertainty relation \cite{Heisenberg_1927,Robertson_1929}: 
\begin{equation}
    (\Delta \hat{A})^2 (\Delta \hat{B})^2\geq \frac{1}{4}\langle i[\hat{A},\hat{B}] \rangle^2 \ .
\end{equation}

\noindent\textbf{Mixed states}.-- Pure states are important since they describe states that are, in some sense, elementary. 
However, for any pure state $\ket{\psi}$, there exists a projective measurement, $\{\hat{\pi}_\psi = \ketbra{\psi}, \hat{\pi}_{\mathrm{not}\psi} = \mathbb{I} - \ketbra{\psi} \}$, with zero variance. This condition is often too stringent, which prompts us to relax it by incorporating into the theory also mixtures of pure states: that is, ensembles. Specifically, mixed states can be decomposed using outer product\footnote{From a physical perspective, outer product of two vectors $\ket{\psi}$ and $\ket{\phi}$, $\ketbra{\psi}{\phi}$, can be thought of as a matrix product of $\ket{\psi}$ and complex conjugated $\bra{\phi}$.} of pure states as per
\begin{equation}
    \hat{\rho} = \sum_i p_i\ketbra{\psi_i} \ ,
    \label{eq:mixed_states}
\end{equation}
where $p_i$ is interpreted as the probability of finding the system in the state $\ket{\psi_i}$. Due to their statistical origin, the description of states using mixed states is alternatively called \emph{density matrices}. To see that mixed states are more general than pure ones, observe that each pure state $\ket{\psi}$ can be written as a density matrix, $\hat{\rho} = \ketbra{\psi}$; however, the converse is not true -- not every state is pure. 

One can also define mixed states mathematically, without any explicit reference to the ensemble:

\begin{definition}
    Mixed state $\hat{\rho}$ is an operator, i.e., a matrix, on the Hilbert space $\mathcal{H}$ such that it is positive semidefinite (PSD)\footnote{That is, all the eigenvalues of $\hat{\rho}$ are real and non-negative.} $\hat{\rho}\succeq 0$ and with normalized trace $\text{Tr}(\hat{\rho}) = 1$.
\end{definition}

\noindent\textbf{Dynamics}.-- Evolution induced on a closed system is given by exponentiation of Hamiltonian (which is an Hermitian operator), $\ket{\psi (t)} = \exp{-itH/\hbar}\ket{\psi}$. 
Such an evolution is necessarily \emph{unitary}, meaning that it preserves the inner product in the Hilbert space: $\braket{\psi(t)} = \braket{\psi}$. 
Alternatively, if we think of states more concretely as vectors $\ket{\psi}$ of dimension $d$, then such a unitary evolution $U$ can be treated as a square matrix of size $d$.  
Therefore, the unitarity condition is rewritten as
\begin{equation}
    \textrm{for all states }\ket{\psi}: \bra{\psi}U^\dagger U \ket{\psi} = \braket{\psi} \Longleftrightarrow  U^\dagger U = \mathbb{I},
\end{equation}
with an analogous $UU^\dagger = \mathbb{I}$.
Consequently, the unitary transformations of quantum states can be represented as unitary matrices.
This might seem trivial since the action of a unitary matrix of size $d$ in a natural way should describe all valid transformations of a state, $\ket{\psi} \mapsto U\ket{\psi}$.\\

\noindent\textbf{Quantum channels}.-- Notwithstanding, the language of unitary operations is proper only for closed systems. 
In some cases, it is a useful approximation; however, no system is truly isolated from the environment. 
If the influence of the outside world is too large then the evolution is no longer unitary, it needs to be described using the language of quantum channels. Beyond dynamics, quantum channels represent the most general physical transformations.  
\begin{definition}
    Channel $\Phi$ is a completely positive trace-preserving\footnote{This condition is added so that the final state is always physical, independently of the initial state.
    Non-completely positive maps can also be physical, provided some initial correlations with the environment~\cite{Carteret_2008}.} (CPTP) mapping from operators in $\mathcal{H}_X$ to operators in $\mathcal{H}_Y$. 
\end{definition}
There are many ways to describe quantum channels~\cite{Bengtsson_2006,Watrous_2018}, but the most common approach involves Kraus operators
\begin{equation}
    \Phi(\hat{\rho}) = \sum_i K_i \hat{\rho} K_i^\dagger,
    \label{eq:kraus_rep}
\end{equation}
where $\sum_i K_i^\dagger K_i = \mathbb{I}$. By construction, Eq.~\eqref{eq:kraus_rep}, the map $\Phi$ is \textit{positive}, that is for all PSD $\hat{\rho}$, $\Phi(\hat{\rho})\succeq 0$. It is not so direct that $\Phi$ according to Eq~\eqref{eq:kraus_rep} is also \textit{completely positive}, i.e., the extended map $\mathrm{Id}\otimes \Phi$, where $\mathrm{Id}$ is the identity map, is positive. Surprisingly, complete positivity is easier to verify than positivity alone. Indeed, thanks to Choi's theorem \cite{Choi_1975}:

\begin{theorem}
\label{th:choi}
    A map $\Phi$ is completely positive if and only if the matrix,

    \begin{equation}
        M_{\Phi} = \sum_{i,j} \ketbra{i}{j}\otimes \Phi(\ketbra{i}{j}) \ ,
        \label{eq:choi}
    \end{equation}
    where $\{\ket{i}\}$ is an orthonormal basis of $\mathcal{H}_X$, 
     is positive semidefinite. 
\end{theorem}

Note that the Choi matrix, $M_{\Phi}$, as defined earlier can be written as \\ \mbox{$M_{\Phi} = (\mathrm{Id}\otimes\Phi)(\ketbra{\phi^+})$}, where $\ket{\phi^+}=\sum_{i=1}^d\ket{i}\ket{i}$ is a maximally entangled state.

\subsection{Bipartite systems}
Sometimes it is useful to divide the Hilbert space of interest into two subsystems A and B, $\mathcal{H} = \mathcal{H}_A \otimes \mathcal{H}_B$. 
This is especially important when we have a clear division between two physical parts of the entire system.\footnote{Nonetheless, in this chapter, we are not concerned with physical constraints that might govern a specific application of mathematical theory. 
Rather, we shall follow the path of introducing all the relevant notions for the general case.} With this picture in mind, any matrix $M$ acting in $\mathcal {H}$ can be interpreted as $M = \sum_i A_i\otimes B_i$ where $\{ A_i \}$ acts in $\mathcal{H}_A$ and $\{B_i \}$ acts in $\mathcal{H}_B$. \\

% \noindent \textbf{Partialology.--} 
For future use, we introduce a couple of concepts that will become exceptionally useful for studying entanglement.

\begin{definition}
    Partial trace over subsystem $B$ of a bipartite matrix $M = \sum_i A_i \otimes B_i$ of order $d_A d_B$ is a matrix $\text{Tr}_B M = \sum_i\text{Tr}(B_i) A_i$ of order $d_A$ (analogously for $\text{Tr}_A$). 
\end{definition}
The usefulness of the above definition in quantum information stems from the fact that if we are interested in only subsystem $A$, state $\hat{\rho}$ behaves exactly as would $\hat{\rho}_A = \text{Tr}_B(\rho)$. 
Therefore, we conclude that partial trace encapsulates all information about the state that is accessible from its constituents. 

\begin{definition}
    Partial transposition over subsystem $B$ of a bipartite matrix $M = \sum_i A_i \otimes B_i$ is $M^{T_B} = \sum_i A_i\otimes B_i^T$, where $T$ is the usual transposition. 
\end{definition}
Such a concept will become useful for entanglement detection through the celebrated Peres-Horodecki criterion (PPT), which will be addressed in the next subsection.  \\

\noindent \textbf{Quantum entanglement}.-- If the subsystems did not interact in the past in a meaningful way,\footnote{For example, one particle on Earth and the other on Mars.} we can reasonably believe that they are not correlated. In such a case, the properties of each system are statistically independent. In particular, we have the factorization of joint observables,  $\langle \hat{O}_A\hat{O}_B'\rangle = \langle \hat{O}_A\rangle \langle \hat{O}_B'\rangle $ where $\hat{O}_A, \hat{O}_B'$ are acting on subsystems $A$ and $B$ respectively.\footnote{That is, $\hat{O}_A = \hat{O}\otimes\mathbb{I}_B$, $\hat{O}_B = \mathbb{I}_A\otimes \hat{O}$.}
If the underlying quantum state is pure, then it is of product form:

\begin{definition}
    Pure state $\ket{\psi}\in\mathcal{H}_A \otimes \mathcal{H}_B$ is called separable when $\ket{\psi} = \ket{\psi_A}\otimes \ket{\psi_B}$. 
\end{definition}
Even if some correlation is exhibited  $\langle \hat{O}_A\hat{O}_B'\rangle - \langle \hat{O}_A\rangle \langle \hat{O}_B'\rangle \neq 0 $, it does not necessarily mean that the underlying state is entangled.  The classical world is full of correlations which are described without invoking the Hilbert space formalism introduced in this chapter. Such cases are addressed with separable mixed states:

\begin{definition}
      A mixed state $\hat{\rho}$ is separable if and only if it admits a decomposition into pure separable states 
      \begin{equation}
          \hat{\rho} = \sum_i p_i \ketbra{\psi_i^A}\otimes\ketbra{\psi_i^B} \ ,
      \end{equation} 
   with $\{ p_i\}$ a probability distribution~\cite{Horodecki_1997}. 
\end{definition}

It is crucial that according to the previous definition, not all separable states are of product form $\hat{\rho} = \hat{\rho}_A \otimes\hat{\rho}_B$, as we allow for statistical mixtures. 

The operational interpretation of such classical correlations is that parties $A$ and $B$ have access to a hidden shared ``coin'' sampling from probability distribution $\{p_i\}$.\footnote{Hidden in the sense that we do not have access to the sampled outcomes.} When outcome $i$ is produced $A$ prepares $\ket{\psi^A_i}$, while $B$ prepares $\ket{\psi^B_i}$ in their part.  
States that cannot be created using the above procedure, i.e., non-separable, are called entangled.\footnote{Observe that here we distinguish between particles A and B. Yet another subset of entanglement study is concerned with \emph{indistinguishable} particles~\cite{Wiseman_2003,Dowling_2006} (see Section~\ref{sec:4}).}
\begin{definition}
    State $\hat{\rho}$ is entangled if it is not separable. 
\end{definition}
Entangled states represent rather strong correlations, in which no classical probability distribution $\{p_i \}$ exists. Along this line, the task of entanglement detection, which is precisely the topic of the next subsection, can be phrased as the task of telling apart quantum from classical correlations.  It is not difficult to show that non-classical correlations, explained by entangled states necessarily involve correlation between incompatible (non-commuting) observables. 

\begin{example}{Understanding quantum entanglement with the spin singlet}
We will settle the concepts defined above with an example. Imagine that you are given the following states in a two-qubit system:   
\begin{align}
\ket{\psi^-} &= \frac{1}{\sqrt{2}}(\ket{+1}\otimes\ket{-1}-\ket{-1}\otimes \ket{+1})\\
\hat{\rho} &= \frac{1}{2}(\ketbra{+1}\otimes \ketbra{-1}+\ketbra{-1}\otimes\ketbra{+1})\ ,
\end{align}
and the spin observables with components in the basis $\{\ket{+1},\ket{-1} \}$:
\begin{equation}
    \hat{\sigma}_x = \begin{pmatrix}
        0 & 1 \\
        1 & 0 
    \end{pmatrix} \ , \ \ 
        \hat{\sigma}_y = \begin{pmatrix}
        0 & -i \\
        i & 0 
    \end{pmatrix} \ , \ \ 
    \hat{\sigma}_z = \begin{pmatrix}
        1 & 0 \\
        0 & -1 
    \end{pmatrix} \ . \ \ 
\end{equation}

Both states $\ket{\psi^-}, \hat{\rho} $ exhibit perfect anticorrelation with respect to $z$, 
\begin{equation}
    C_{zz} = \langle \hat{\sigma}_z\otimes\hat{\sigma}_z \rangle - \langle \hat{\sigma}_z \otimes \mathbb{I}\rangle\langle  \mathbb{I}\otimes \hat{\sigma}_z\rangle = -1
\end{equation}
However, it turns out that $\ket{\psi^-}$ is entangled while $\hat{\rho}$ is separable.  So, what is the difference between these two states?

The key observation is that $\ket{\psi^-}$ is also perfectly anticorrelated with respect to $x$, $y$, and, in general -- because of the rotation invariance -- for any linear combination, $n_x\hat{\sigma}_x + n_y\hat{\sigma}_y + n_z\hat{\sigma}_z$. A possible classical explanation would be that A and B have both access to a shared ``coin'' that determines A to prepare a state with well-defined spin in all three components ``$\ket{a_xb_yc_z}$'' , $a,b,c\in \{\pm1\}$ and $B$ the opposite ``$\ket{-a_x-b_y-c_z}$''. However, the spin components $x,y,z$ are incompatible (their corresponding operators do not commute), so the state ``$\ket{a_xb_yc_z}$'' does not exist as not all three components can be well-defined simultaneously. In such a case, quantum entanglement is necessary to explain these correlations.

% It is of utmost importance to notice that according to the previous definition not all separable states are of product form $\hat{\rho} = \hat{\rho}_A \otimes\hat{\rho}_B$, as we allow for mixtures. For example:
%      \begin{equation}
%         \hat{\rho} = \frac{1}{2}\left(\ket{0}\bra{0}\otimes\ket{0}\bra{0} + \ket{1}\bra{1}\otimes\ket{1}\bra{1}\right)
%     \end{equation}
%     is separable but not a product state. When measuring one of the subsystems in the computational basis $\{\ket{0}, \ket{1}\}$, one obtains perfectly correlated outcomes, either all subsystems are in the state $\ket{0}$ or $\ket{1}$. The state is separable and not entangled despite the correlations because one can generate this state from classical operations alone. 
\end{example}

\subsection{Detection of entanglement}

One of the main topics of quantum information is the detection and quantification of entanglement. The first question that we raise is concerning the discrimination of entanglement:

\begin{svgraybox}
    Given a bipartite $\hat{\rho}$ acting in $\mathcal{H}_A\otimes\mathcal{H}_B$, is it separable or entangled?
\end{svgraybox}

\noindent As simple as it is posed, deciding whether a state is entangled or separable is NP-hard~\cite{Gurvits_2003}. The challenge concerns the mixed states. Conversely, for pure states, it is solved in a rather simpler way.

\subsubsection{Pure states}

In the case of bipartite pure states, this is usually done based on a simple observation concerning partial trace. 

\begin{overview}{Purity condition}
Separable pure state $\hat{\rho} = \ketbra{\psi_A}\otimes \ketbra{\psi_B}$ is pure after partial trace, $\text{Tr}_B \hat{\rho} = \ketbra{\psi_A}$. 
Conversely, an entangled pure state is not pure after a partial trace.
    
\end{overview}

\noindent \textbf{Entropic quantities}.-- To connect it with entanglement, we need to quantify how pure the state is. 
\begin{definition}
    The von Neumann entropy $S$ of state $\hat{\rho}$ is defined as $S(\hat{\rho}) = - \text{Tr}\,(\hat{\rho}\ln \hat{\rho})$, with $\ln$ being matrix logarithm. 
\end{definition}
To see that this is a valid quantifier, note that pure states have von Neumann entropy 0, since in the eigenbasis all elements are either 0, apart from the one that equals unity. 
The other extreme case is the maximally mixed state, defined as $\mathbb{I}/d$, where $d$ is the dimension of the Hilbert space. 
It is maximally mixed in the sense that its von Neumann entropy is maximal and equal to $\ln d$.
% \begin{definition}
%     Purity $\gamma$ of state $\rho$ is defined as trace of the squared state, $\gamma (\rho) = \text{Tr}\,\rho^2$. 
% \end{definition}
% To see that this is a valid quantifier, note that pure states have purity 1, since their density matrix is a projection, $(\ket{\psi}\bra{\psi})^2 = \ket{\psi}\bra{\psi}$.
% The other extreme case is the maximally mixed state, defined as $\mathbb{I}/d$, where $d$ is the dimension of the Hilbert space. 
% It is maximally mixed in the sense that its purity is minimal, $\text{Tr}\,\,\mathbb{I}/d^2 = 1/d$.

Using the above observation concerning partial trace, we can connect the level of purity (or, conversely, its von Neumann entropy) of the partial state with the entanglement of the full state.  

\begin{definition}\label{def:entropy}
    The entropy of entanglement $E$ of a bipartite state $\hat{\rho}$ is defined as the von Neumann entropy of either of its reduced states, $E (\hat{\rho}) = S(\hat{\rho}_A) = S(\hat{\rho}_B)$. 
\end{definition}
It is not hard to prove that although reduced states $\hat{\rho}_A$ and $\hat{\rho}_B$ are \emph{not} equal, their von Neumann entropies are, cf. Ref.~\cite{Nielsen2012}. 
In contrast to the least entangled, separable states (not entangled at all), we now can introduce a maximally entangled state called Bell state $\ket{\phi^+} = \frac{1}{\sqrt{d}}\sum_i^d \ket{i}\ket{i}$. 
Its partial trace is the maximally mixed state, thus it is truly a state of maximal possible entanglement. 
Clearly, acting with \emph{local} unitary channel $U_A\otimes U_B$, we are unable to change the entanglement value of any state $\hat{\rho}$. 
Therefore, instead of one maximally entangled Bell state, we rather have a full family of them, defined as $\frac{1}{\sqrt{d}}\sum_i^d U_A\ket{i}\otimes U_B\ket{i}$.
Furthermore, all of the bipartite states of maximal entropy of entanglement are local unitary equivalent to the Bell state $\ket{\phi^+}$, which encourages us to treat the Bell state as a gold standard for maximal entanglement~\cite{Plenio_2007}. \\

\noindent \textbf{Schmidt decomposition}.-- Another useful notion for studying the entanglement of pure quantum states is the Schmidt decomposition, which is the singular value decomposition (SVD) of the state's coefficients. 

\begin{lemma}
    Every bipartite pure state $\ket{\psi}\in\mathcal{H}_A \otimes \mathcal{H}_B$ can be written using its Schmidt decomposition, i.e., $\ket{\psi} = \sum_i \sqrt{\lambda_i} \ket{\xi_i}\otimes\ket{\phi_i}$, with non-negative $\lambda_i$ forming a probability distribution, while $\ket{\xi_i}\in\mathcal{H}_A$ and $\ket{\phi_i}\in \mathcal{H}_B$.
\end{lemma}
The advantage of the Schmidt decomposition over an arbitrary basis of density matrices in the bipartite Hilbert space $\mathcal{H}_A\otimes \mathcal{H}_B$ is that it requires only one index, thus it is a form of diagonalization. 
Mathematically, the validity of the above lemma is proved using singular value decomposition~\cite{Nielsen2012}. 

The coefficients of the Schmidt decomposition $\boldsymbol{\lambda} = \{\lambda_i\}$ are a useful tool for entanglement classification since one can directly infer the entropy of entanglement, as well as other measures of entanglement such as the Schmidt rank. 
Importantly, in the pure bipartite case, essentially all entanglement measures are equivalent, thus it is enough to consider only one of them (typically the entropy of entanglement)~\cite{Plenio_2007}. 

\subsubsection{Mixed states}

\noindent \textbf{Convex roof construction}.-- The entropic quantifiers defined in the previous paragraph are not useful for detecting entanglement in mixed states. Indeed, take for example the entanglement entropy $E$. Such quantity is able to capture correlations but fails to distinguish between those stemming from entanglement or from classical mixing. For instance, both states $\ket{\phi^+} = (\ket{00} + \ket{11})/\sqrt{2}$ and $\hat{\rho} = (\ketbra{00} + \ketbra{11})/2$ have maximal entropy of reduced states, but $\hat{\rho}$ is separable.   

A rigorous way to extend entanglement measures of pure states into mixed states is via the convex roof construction: 

\begin{overview}{Convex roof construction}
Given an entanglement quantifier $F$ valid for pure states $\ket{\psi}$, it can be extended to mixed states $\hat{\rho}$ by optimizing over all possible decompositions of $\hat{\rho}$ into pure states:   

\begin{equation}
     F(\hat{\rho}) = \min_{\sum_i p_i\ketbra{\psi}_i = \hat{\rho}} \sum_{i}p_i F(\ket{\psi_i})
\end{equation}
    
\end{overview}

\noindent In practice, the previous minimization is never carried out explicitly (which would be impossible), but implicitly via convex optimization techniques \cite{Toth_2015} which allow for some cases the derivation of closed formulas for $F(\hat{\rho})$. \\

\noindent \textbf{PPT criterion}.-- Entanglement discrimination in mixed states is more complicated than its pure counterpart --- one needs to resort to other methods, out of which arguably the simplest is the positive partial transpose (PPT) criterion~\cite{Horodecki_1996, Peres_1996}. 

\begin{overview}{Peres-Horodecki or positive partial transpose (PPT) criterion}
Let $\hat{\rho}$ be a bipartite state acting on $\mathcal{H}_A\otimes\mathcal{H}_B$. 
If $\hat{\rho}$ is separable, then the partial transpose is PSD, $\hat{\rho}^{T_B}\succeq 0$. 
Conversely, if partially transposed $\hat{\rho}^{T_B}$ is not a PSD matrix, i.e., it has at least one negative eigenvalue, then $\hat{\rho}$ is entangled.

%Let us define partial transposition on a bipartite state $\rho = \sum_i A_i \otimes B_i$ as $\rho^{T_B} = \sum_i A_i \otimes B^T_i$, analogously for the subsystem $A$. Then, state $\rho$ is entangled if partially transposed $\rho^{T_B}$ is not a positive matrix, i.e., has at least one negative eigenvalue. 
\end{overview}

\noindent Since $\hat{\rho}^{T_A} = (\hat{\rho}^{T_B})^T$, the criterion does not depend on the choice of the subsystem. 
For small subsystems $A$ and $B$ -- of dimensions $(d_A,d_B) = (2,2)$ or $(2,3)$ -- this condition is also sufficient for entanglement. 

However, for higher-dimensional systems, PPT criterion is only one-way -- there are entangled states with PSD partial transpose (so-called PPT entangled states)~\cite{Horodecki_1998,Lewenstein_2000,Bruss_2002_2}.\\

\noindent\textbf{Detection methods}.-- Here, we list well-established methods for entanglement detection, both bipartite and multipartite: entanglement witnesses (see subsequent paragraphs), Bell inequalities (see Section~\ref{subsec:Bell_nonlocality}), matrix realignment~\cite{Chen_2003,Rudolph_2005}, nonlinear properties of more than one copy of a state~\cite{Horodecki_2003}, quantum switch-aided protocol~\cite{Abanin_2012}, Gilbert's algorithm~\cite{Shang_2018,Pandya_2020,Wiesniak_2020}, Lewenstein-Sanpera decomposition~\cite{Lewenstein_1998}, trace polynomial inequalities~\cite{Rico_2023}, moments of many-body systems~\cite{Gessner2019,Frerot_2022,Frerot_2021}, quantum Fisher information~\cite{Hauke_2016,Mathew_2020,Scheie_2021,MullerRigat2023}, randomized measurements~\cite{Elben_2020,Vermersch2023}, and spin-squeezing inequalities~\cite{Toth_2009,Vitagliano2014,MullerRigat2022}. 
The detection of entanglement is an active field also from the experimental perspective~\cite{Brukner_2006,Esteve_2008,Haas_2014,Lanting_2014}.

\begin{overview}{Lewenstein-Sanpera decomposition}
    Any bipartite density matrix $\hat{\rho}\in \mathbb{C}^d\otimes\mathbb{C}^d$ can be decomposed as per
    \begin{equation}
        \hat{\rho}=\lambda \hat{\rho}_s +(1 - \lambda)\hat{P}_e\ ,
    \end{equation}
where $\lambda \in [0,1]$, $\hat{\rho}_s$ is a separable state and $\hat{P}_e = \ketbra{\psi}$ is a rank-1 projector onto an entangled state $\ket{\psi}$.

The separable state $\hat{\rho}_s$ such that $\lambda$ is maximal constitutes the \emph{best separable approximation} of $\hat{\rho}$. In particular, if the optimal  $\lambda = 1$, we deduce that $\hat{\rho} = \hat{\rho}_s$ is separable. 
Finally, such a method works for any convex subset of states, beyond the separable set. 
\end{overview}

\noindent \textbf{Entanglement and thermodynamics.--}
Usually entanglement is observed only in microscopic systems. Nonetheless, for specific choices of the Hamiltonian, entanglement can be related to several macroscopic thermodynamical properties, such as temperature~\cite{Nielsen_1998}, magnetic susceptibility~\cite{Wiesniak_2005,Brukner_2006}, or heat capacity~\cite{Wiesniak_2008}, that can act as an entanglement witnesses.

In many systems, the ground state is entangled and so are states with low energy. 
The entanglement gap for this case is defined as the energy difference between the ground state and the minimal energy attainable by separable states.
Provided the existence of the entanglement gap, energy can act as an entanglement witness for thermal states of Hamiltonians with entangled ground state~\cite{Dowling_2004,Toth_2005,Tura_2017}. Similarly, internal energy outside of the separable bounds indicates the presence of entanglement even for states out of equilibrium~\cite{Guhne_2005}.\\

For the purposes of this chapter, we shall not elaborate further on the topic of detection of entanglement as excellent positions already exist, such as the review by G\"{u}hne and T{\'{o}}th~\cite{Guhne_2009}. 
Let us now focus on entanglement witness and linear maps methods.\\ 

\subsubsection{Entanglement witnesses}

\begin{figure}
    \centering
    \includegraphics[scale=0.3]{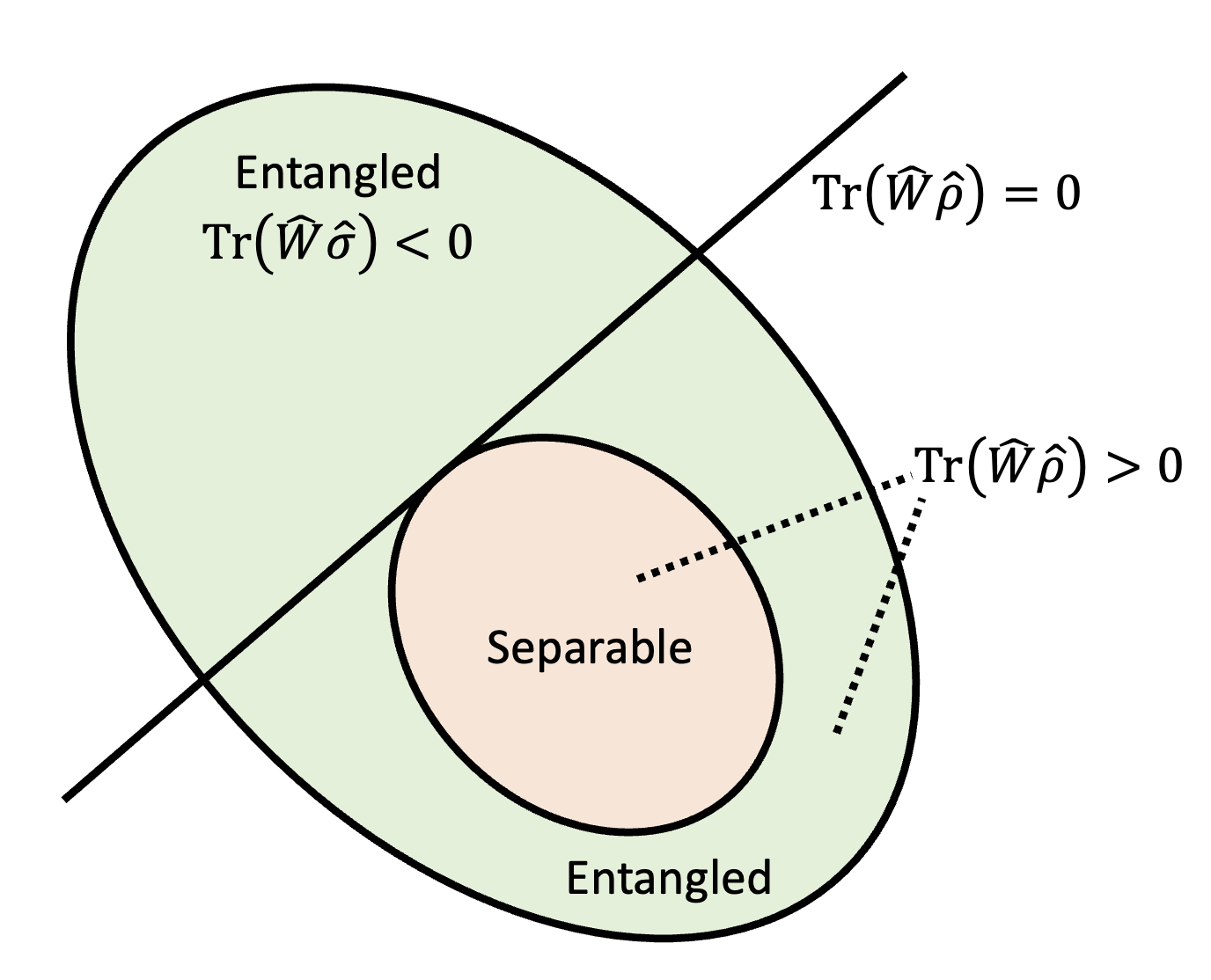}
    \caption{Entanglement witness $W$ divides the set of states into those which are entangled ($\text{Tr}(\hat{W}\hat{\rho}) < 0$) and the rest, which might be entangled or not ($\text{Tr}(\hat{W}\hat{\rho}) \geq 0$).
    We say $\hat{W}$ is optimal ~\cite{Lewenstein_2000} if there exist separable state $\hat{\rho}_S$ on the boundary, i.e., $\text{Tr}(\hat{W}\hat{\rho}_S) = 0$.}
    \label{fig:entanglement_witness}  % GMR: Don't change label as I'm ref this
\end{figure}

% \noindent\textbf{Entanglement witnesses}.--  
As such, the PPT is the most basic scheme for entanglement quantification; nevertheless, typically, one needs other criteria if this test is inconclusive. 
To this end, special operators have been introduced, called witnesses~\cite{Horodecki_1996}.
\begin{definition}
    Entanglement witness $\hat{W}$ is an operator such that for all separable states $\text{Tr}\,(\hat{W}\hat{\rho})\geq 0$ and there exists at least one entangled state $\hat{\sigma}$ for which $\text{Tr}\,(\hat{W}\hat{\sigma})<0$. 
\end{definition} 
Similar to the PPT criterion, the entanglement witness method is not conclusive, for every $\hat{W}$ there are entangled states that are not certified to be entangled, as illustrated by Figure~\ref{fig:entanglement_witness}. 
Nevertheless, for every entangled state $\hat{\sigma}$ it is possible to find $\hat{W}_\sigma$ such that $\text{Tr}\,(\hat{W}_\sigma\, \hat{\sigma})<0$, thus certifying its entanglement~\cite{Pittenger_2002}. 

Entanglement witnesses can arise from experimentally feasible mutually unbiased bases~\cite{Spengler_2012}, while connecting also to quantum state tomography~\cite{Chruscinski_2018,Bae_2019,Hiesmayr_2021,Bavaresco_2018}. 
Another way of constructing them is the semi-definite programming~\cite{Huber_2022}.
Furthermore, apart from a straight division of quantum states into two parts, there are also nonlinear entanglement witnesses~\cite{Horodecki_2003,Guhne_2006}. \\

% After all the previous enumeration of recent theoretical efforts, w
Now, we review an important tool to characterize entanglement witnesses via polynomials.   

\begin{overview}{Entanglement witnesses: a polynomial problem}
The problem at hand is deciding whether an operator $\hat{W}$ is an entanglement witness. If it is so, for any separable state $\hat{\rho}_{\text{SEP}}$, $\mathrm{Tr}(\hat{W}\hat{\rho}_{\text{SEP}})\geq 0$. By convexity, it is sufficient to prove such condition for pure states $\ket{\Psi} = \ket{\psi_A}\otimes \ket{\psi_B}$. Let $\{a_i = \bra{i}\ket{\psi_A}, \,b_i = \bra{i}\ket{\psi_B} \}$ be the components of the local vectors in an orthonormal basis. Then the positivity condition becomes:

\begin{equation}
\bra{\Psi}\hat{W}\ket{\Psi} = \sum_{i,j,k,l}W_{ij,kl}a_ib_ja^*_kb^*_l\geq 0 \ ,
\end{equation}
where $W_{ij,kl}$ are the coordinates of $\hat{W}$ with respect to the local basis. 

We conclude that the expectation value is non-negative if and only if the quartic polynomial $p_4 = \sum_{i,j,k,l}W_{ij,kl}a_ib_ja^*_kb^*_l$ is non-negative in the complex domain.

\end{overview}

\noindent \textbf{Sum-of-squares and decomposable entanglement witnesses.--} A certain class of polynomials that are positive are those that can be written as a sum of squares (SOS) of other polynomials of lower degree, $p_4 = \sum_{i}|p_2^{(i)}|^2$. Such class is feasible to characterize via convex hierarchies \cite{Lasserre2004}, and gives rise to entanglement witnesses of the form $\hat{W} = \hat{P} + \hat{Q}^{T_B}$ for $\hat{P},\hat{Q}\succeq 0$, which are called \textit{decomposable}. As proven in Refs.~\cite{Woronowicz1976,Lewenstein_2000}, 

\begin{theorem}
Decomposable witnesses are equivalent to the PPT condition. That is, all entangled states detected by witnesses of the form $\hat{W} = \hat{P} + \hat{Q}^{T_B}$ for $\hat{P},\hat{Q}\succeq 0$ are not PPT. The converse is also true: for any non-PPT state, there is a decomposable witness detecting it.  
\end{theorem}
Let us consider the converse statement: are all positive polynomials SOS? Hilbert proved that in general this is \emph{not} the case \cite{Hilbert1888}. Decades later, the first counterexample was found by Motzkin~\cite{Motzkin_1967}.\footnote{It reads $p(x,y) = x^4y^2 + x^2y^4+1-3x^2y^2$ over the real field.}
It is remarkable that such negative result has connections to the matter of this chapter. In our language, it implies that PPT is not sufficient, there exist PPT entangled states and, in general, that entanglement detection is hard (more precisely, NP-hard).

\begin{comment}
\subsection{Quantification of entanglement}
Let us now extend the study from the previous subsection into quantifying the entanglement.
This is a more demanding task than just deciding whether the state is entangled or not, especially given that quantifying entanglement is an NP-hard problem~\cite{Gharibian_2010}.

In order to quantify the entanglement, we need to define a measure of it, which should satisfy the following criteria
\begin{enumerate}
    \item 
\end{enumerate}

Axioms of entanglement measure.

Pure states at first...

Another way to quantify entanglement of a given state is to assess its usefulness in practical tasks.
To operational entanglement measures one can count distillable entanglement, entanglement cost, distillable key rate~\cite{Plenio_2007}.

\begin{svgraybox}
Convex-roof construction is a way to extend constructions valid for pure states into all mixed states~\cite{?}.
\end{svgraybox}
\end{comment}

\subsubsection{Linear maps}

%Environment, Kraus representation (maybe), citation to~\cite{Watrous_2018,Bengtsson_2006} CPTP maps -- definition, LOCC via Kraus and POVMs, separable operations as $\rho \mapsto \sum_i \big(A_i\otimes B_i\big) \rho \big(A^{\dagger}_i\otimes B^{\dagger}_i\big)$. The next subsection explores the usefulness of quantum entanglement. 

Until now, we have developed the theory of entanglement detection and presented some of the most well-known strategies such as PPT and entanglement witnesses. Here, we will provide stronger criteria based on linear maps~\cite{Wolf_2012}.    

We divide this subsection into necessary and sufficient conditions for separability. The first one corresponds to witnesses, which rule out a separable description. Verifying that a state is \emph{not} separable is much less challenging than proving that it actually \textit{is}.  
In the second part, we address this latter problem via sufficient separability criteria.  \\

% \subsubsection{Necessary criteria for separability}

\noindent \textbf{Necessary criteria for separability.--} Maps that are positive but not completely positive (PnCP) are generally not physical, but they are instrumental for entanglement detection. 

\begin{overview}{Positive but not completely positive (PnCP) maps}
Consider a bipartite state $\hat{\rho}$ acting on $\mathcal{H}_A\otimes\mathcal{H}_B $ and a positive map $\mathcal{E}$ transforming operators on $\mathcal{H}_B$. Then,  

\begin{equation}
    (\mathrm{Id}_A\otimes \mathcal{E})(\hat{\rho})\nsucceq 0 \Longrightarrow \hat{\rho}\ \mathrm{is\  entangled}
    \label{eq:PnCP_criteria}
\end{equation}    
\end{overview}
\noindent Of course, if map $\mathcal{E}$ is CP, this cannot be used. The condition above generalizes the PPT criteria as presented earlier from the transposition to an arbitrary positive map.\\

There is a correspondence between maps and entanglement witnesses via the Choi matrix formalism that we introduced in Eq.~\eqref{eq:choi}. Specifically, the Choi matrix of the dual map $\mathcal{E}^\dagger$,\footnote{The dual of the map $\mathcal{E}$ is defined as the map $\mathcal{E}^\dagger$ such that $\mathrm{Tr}[\hat{\rho}\mathcal{E}(\hat{\sigma})] = \mathrm{Tr}[\mathcal{E}^\dagger(\hat{\rho})\hat{\sigma}] $ for any pair of matrices $\hat{\rho},\hat{\sigma}$.}  $ \hat{M}_{\mathcal{E}^\dagger} = (\mathrm{Id}_A\otimes\mathcal{E}^\dagger)(\ketbra{\phi^+}) $ is an entanglement witness:  

\begin{equation}
   \mathrm{Tr}(\hat{M}_{\mathcal{E}^\dagger}\hat{\rho})<0 \Longrightarrow \hat{\rho} \ \mbox{is\ entangled} \ .
   \label{eq:choiwitness}
\end{equation}
Note that, if $\mathcal{E}$ is CP, $\mathcal{E}^\dagger$ so is. In such case, the corresponding entanglement witness is trivial, as, according to Theorem~\ref{th:choi}, $\hat{M}_{\mathcal{E}^\dagger}$ would be PSD.

Evidently, the condition on maps Eq.~\eqref{eq:PnCP_criteria} is stronger than a concrete entanglement witness realization Eq.~\eqref{eq:choiwitness}. However, witnesses can be implemented in the experiments, while PnCP maps do not have a physical realization. \\

% \subsubsection{Sufficient criteria for separability}

\noindent \textbf{Sufficient criteria for separability.--} Although lesser-known, linear maps can be employed for the complementary problem to derive \emph{sufficient} conditions for separability beyond the PPT condition for low dimensionalities. In Ref.~\cite{Lewenstein2016}  new techniques in this regard were introduced, which were extended to the multipartite setting in Ref.~\cite{Tura2018}. 
Here, we offer an example of the proposed methodology.

\begin{example}{Inverting linear maps to detect separable states}
Let us consider a bipartite system $\mathbb{C}^2\otimes\mathbb{C}^d$ and $\Lambda_\alpha$ be a family of reduction maps parametrized by $\alpha\in \mathbb{R}$, defined as: 
\begin{equation}
    \Lambda_\alpha(\hat{\varrho}) = \mathrm{Tr}(\hat{\varrho})\mathbb{I} +\alpha\hat{\varrho} \ .
    \label{eq:red_map}
\end{equation}

One can show that:

\begin{itemize}
    \item For the range $\alpha\in  [-1,2]:= R $ the transformation maps any state to a separable state: 
    \begin{equation}
        \forall \hat{\varrho}\succeq 0,\ \Lambda_{\alpha\in R}(\hat{\rho})\ \mbox{is\ separable} \ .
    \end{equation}
    The lower bound is already deduced by imposing positivity of the map. For the upper bound, see Ref.~\cite{Lewenstein2016}. 
    
    \item The map can be readily inverted and its inverse reads:   $\Lambda^{-1}_\alpha(\hat{\sigma})= (\hat{\sigma} - {\rm
Tr}(\hat{\sigma})\mathbb{I}/(2d+\alpha))/\alpha$.
\end{itemize}

These two observations allows us to assert $\Lambda^{-1}_{\alpha\in R}(\hat{\sigma}) \succeq 0 \Longrightarrow \hat{\sigma}\ \mbox{is\ separable}$. Applying this finding for the extreme case $\alpha = 2$, yields the following criterion: 
\begin{align}
  \hat{\sigma} - \frac{\mathbb{I}}{2d+2}\geq 0\Longrightarrow \hat{\sigma}\mathrm{~is~separable}
\end{align}
Such criterion detects, e.g, states in the separable ball around the completely mixed state, $\mathrm{Tr}(\hat{\rho}^2)\leq 1/(2d-1) $.
    
\end{example}

By definition, quantum entanglement is intrinsically linked to a Hilbert space. However, there exists a deeper notion of non-classicality, called Bell nonlocality, without explicit reference to such mathematical construction. In the next section, we give a pedagogical introduction to the field, which will be later developed in the last part of this chapter.

\subsection{Bell nonlocality}\label{subsec:Bell_nonlocality}
The inherent unpredictability of quantum mechanics is so different from the everyday world that even some of those who laid the groundwork for the formulation of quantum theory could not fully believe in it.
Einstein, Podolsky, and Rosen in 1935 claimed that this unpredictability is a feature arising due to its incompleteness -- which was later dubbed as EPR paradox~\cite{Einstein_1935}. 
In a nutshell, they argued that since entanglement can affect far objects in an instant, there must exist an underlying theory that is able to correctly attribute characteristics to a state of the system, such that the predictions of quantum mechanics are recoverable. 

Those arguments were refuted by others~\cite{Bohr_1935}; however, for many years it was only a philosophical debate with physical implications nowhere to be found. 
Only in 1964, John Bell has proven that indeed, quantum mechanical unpredictability can be tested~\cite{Bell_1964}. 
What was later named ``Bell inequalities'' are outcomes of certain measurements that must be confined to a specific region if we assume the incompleteness of quantum mechanics.\footnote{To be more precise, such requirement is called ``local realism''.}

\begin{figure}
    \centering
    \includegraphics[scale=0.9]{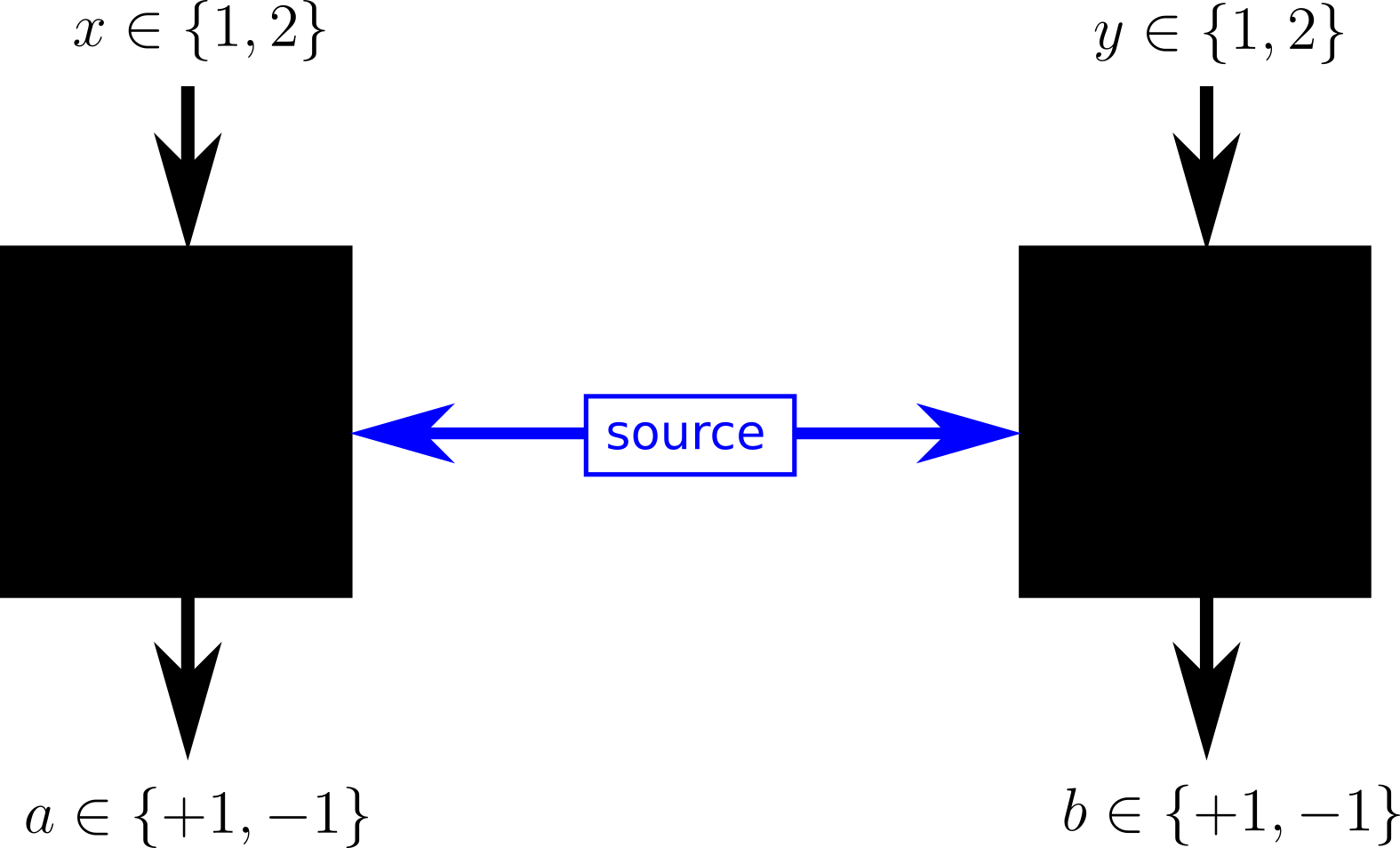}
    \caption{Both parties to the CHSH game receive half of the entangled state from the source, which allows them to break the Bell inequality on the outputs $a$ and $b$.}
    \label{fig:Bell_scenario}  
\end{figure}

Here, we shall discuss the most famous Bell inequality introduced by Clauser, Horne, Shimony, and Holt (CHSH)~\cite{Clauser_1969}. 
Two players, having access to a source of entanglement, are independently given one of the two possible inputs\footnote{Drawn uniformly and randomly.} $x \in \{1,2\}$ to Alice and $y \in \{1,2\}$ to Bob, as in Figure~\ref{fig:Bell_scenario}. 
Then, they want to maximize the following sum of expectation values of four observables $A_i$ and $B_j$
\begin{equation}
    CHSH = \langle A_1 B_1 \rangle + \langle A_1 B_2 \rangle + \langle A_2 B_1 \rangle - \langle A_2 B_2 \rangle,
\end{equation}
while the observables can take two values $\{+1,-1\}$. 

Such value is bounded from above by 2 if the underlying theory is classical, i.e., the values of operators were chosen but are unknown to us. 
However, for maximally entangled Bell states of two qubits, it is possible to achieve theoretically $CHSH = 2\sqrt{2}$. 

It took many years to achieve an experimental confirmation of this breaking of the Bell inequality; however, eventually, those efforts to disprove classical theory underlying the quantum one were awarded the Nobel prize in 2022 ``for experiments with entangled photons, establishing the violation of Bell inequalities and pioneering quantum information science''~\cite{NobelPrize_2022}.

The utility of Bell nonlocality and inequalities arising from it are abundant in quantum information; they can be used in generation as well as certification of true randomness, inaccessible by anyone else than the current user of a device~\cite{Pironio_2010}.
Moreover, such inequalities can be used as entanglement witnesses~\cite{Terhal_2000}.
In the following Section~\ref{subsec:Bell_correlation} we shall elaborate more on the applications of Bell inequalities in the many-body setting. \\

Without a doubt, entanglement is one of the cornerstones of quantum mechanics.
We close the section by summarizing some of the applications in which quantum entanglement (beyond classical correlations) plays a crucial role.

\subsection{Applications of entanglement}

Apart from its conceptual importance, entanglement is a necessary prerequisite for various quantum information protocols. 
A plethora of them are achievable only due to entanglement, e.g., quantum key distribution~\cite{Wiesner_1983,Ekert_1991}, entanglement catalysis~\cite{Jonathan_1999}, quantum teleportation~\cite{Bennett_1993}, quantum secret sharing~\cite{Cleve_1999}, one-step quantum secure direct communication~\cite{Sheng2022}, quantum repeaters~\cite{Briegel_1998}, state preparation~\cite{Mora_2005}, improvement of sensitivity of measurements~\cite{SchleierSmith_2010,Gross_2010,Riedel_2010}, and enhanced quantum machine learning~\cite{Sheng2017, dawid2022modern}.
These setups generally require maximal (or at least sufficiently high) entanglement in order to provide an improvement over their classical counterparts. \\

% \begin{svgraybox}
%     Entanglement distillation and bound entanglement
% \end{svgraybox}

\noindent \textbf{Entanglement distillation.--} Therefore, we also need a way to refine entanglement from many states of lower-quality entanglement via a process called entanglement distillation (purification)~\cite{Bennett_1996,Bennett_1996_v2,Deutsch_1996}. 
Distillation has been proposed~\cite{Pan_2001,Sheng_2008,Zwerger_2013,Zwerger_2014,Yan_2021,Qin_2023} and performed~\cite{Kwiat_2001,Pan_2003,Zhao_2003,Yamamoto_2003,Walther_2005,Reichle_2006,Kalb_2017,Hu_2021,Ecker_2021,Yan_2022} in laboratory settings, also in the multipartite scenario~\cite{Murao_1998,Sheng_2009}. \\

% \begin{svgraybox}
    
% \end{svgraybox}

\noindent \textbf{Bound entanglement.--} Although entanglement distillation is a powerful technique, not all entangled states can be converted to a maximally entangled state, even with an arbitrary number of copies. 
We call such states \emph{bound} entangled~\cite{Horodecki_1998}.
As an example of such states, all PPT states are undistillable.\footnote{In fact, one of the biggest open questions of entanglement theory concerns the reverse problem -- are all bound entangled states also of positive partial transpose~\cite{Horodecki_2022}?}

Even though bound entanglement is not useful at distilling maximally entangled states, it can still be used for 
entanglement activation~\cite{Horodecki_1999}, metrology~\cite{Pal_2021}, quantum steering~\cite{Moroder_2014}, distillation of secure quantum keys~\cite{Horodecki_2005,Horodecki_2008}, and for teleportation~\cite{Masanes_2006}.
Therefore, in analogy to ``normal'' entanglement witnesses, there exist some methods to witness bound entanglement~\cite{Horodecki_1999,Lewenstein_2000,Rudolph_2003,Hiesmayr_2013}.
Another way of detecting bound entanglement involves using mutually unbiased bases~\cite{Bae_2022}, see the references therein for construction of bound entangled families and more about their detection.\\

Having explained the simplicity of the bipartite setting, let us now move on to exploring the richness of multipartite entanglement.

% \begin{svgraybox}
%     Other applications
% \end{svgraybox}

% \subsection{Conceptual topics related to entanglement}
% Bell's theorem, Hidden variables, GYNI games

% \textbf{Basics of QI (states, channels, Unitary, non-unitary, measurements\grm{Actually, POVMs are probably not needed.}})
% Bipartite systems (Definitions, theorems, Schmidt decomposition, Horodecki business, PPT criterion, Entanglement witnesses, Choi isomorphism\grm{not sure whether this isomorphism is needed at all}, Separability criterion,  Maximally entangled states, LOCC/SLOCC classification, Entanglement and channels (entangling power))
% Nonlocality and entanglement (nonlocality without entanglement, GYNI games, Hidden variable theory and Bell's theorem, contextuality), physical examples of qudits, like maybe Cooper pairs. Entanglement swapping

% \grm{Entanglement and channels}

% \guillem{Nonlocality and entanglement (nonlocality without entanglement, GYNI games, Hidden variable theory and Bell's theorem, contextuality), physical examples of qudits, like maybe Cooper pairs. I would not enter too much in the foundations business, because we won't need it for the main part (many-body)}

% \include{3_multipartite (Anubhav)/3_multipartite}

\section{Entanglement in multipartite systems}\label{sec:3}

\textit{There is more to this world than just two particles.}
In the previous sections, we have already discussed the concept of entanglement in quantum theory and its implications for the various physical concepts. As of now, we have only considered the simplest case of bipartite entanglement (entanglement between two subsystems), which automatically raises the question of whether we can extend this notion to entanglement between many subsystems. The answer is yes, and we refer to it as the multipartite entanglement. However, the concept of multipartite entanglement is not as simple and straightforward as bipartite entanglement, which has led to an extensive study of quantification and characterization of properties of multipartite entanglement over the past three decades~\cite{Vidal_2000, Bennett_2000, Bruss_2002, Bengtsson_2006, Plenio_2007, Horodecki_2009, Augusiak_2012, Bertlmann2023}.

The theory of multipartite entanglement is richer and much more extensive than the bipartite case as there are exponentially many number of ways in which an N-partite system can be partitioned; each partition giving rise to a different set of locality constraints. 
In the previous section, we have presented various ways for detection of bipartite entanglement and separability. 
Nonetheless, multipartite entanglement can be present even if the state is separable with respect to every bipartition~\cite{Bennett_1999_2}.
Due to those reasons, it is difficult to characterize and come up with a single theory of multipartite entanglement, rather it depends on the resources and tasks in hand. 
In this section, we will start by defining the structure of states in the multipartite setting, and then discuss the elaborate resource theory of entanglement and various invariants and entanglement measures that allow us to classify the multipartite entanglement classes. We apply this discussion to the example of three qubits and study the various entanglement classes and structures that come out of it.

\subsection{To be \textit{separable} or not to be \textit{separable}}

Starting with pure states, an unentangled state vector in a multipartite Hilbert space $\mathcal{H} = \mathcal{H}_1 \otimes \mathcal{H}_2 \otimes ... \otimes \mathcal{H}_N$ of $N$-distinguishable constituents, is given by the product state of the form
\begin{equation}\label{Eq:pure_fullseparable}
    \ket{\Psi} = \ket{\psi_1} \otimes \ket{\psi_2} \otimes ... \otimes \ket{\psi_N}.
\end{equation}
Any state vector which is \emph{not} of this form is said to be entangled.\\

Unlike the case of bipartite systems, there are many ways in which one can create partitions in the multipartite system over the different parties, which gives rise to the notion of partial separability of a state based on these partitions.\\

\noindent\textbf{Partial separability in pure states}.-- The Hilbert space of $N$-parties can be coarse-grained by creating partitions $\{I_1, I_2,..., I_k\}$ such that $\cup_i^k I_i = I$ where $\{I_l\}$ are the disjoint subsets of indices, with $I = \{1,2,3,...,N\}$. Due to this, a state that may have been previously entangled will now be separable under such partitions. This is called partial separability and a state which is partially separable under such partitions is expressed as
\begin{equation}
    \ket{\Psi} = \ket{\psi_{I_1}}\otimes \ket{\psi_{I_2}} \otimes ... \otimes \ket{\psi_{I_k}},
\end{equation}
which looks similar to the state from Eq.~\eqref{Eq:pure_fullseparable}, with the only difference that $\ket{\psi_{I_i}}$ is a collective pure state of the parties belonging to the subset $I_i$.\\

Based on this notion of partial separability, there exist states which are the multipartite generalization of the maximally entangled bipartite states introduced in the previous section, called the absolutely maximally entangled (AME) states.

\begin{example}{Absolutely maximally entangled (AME) states}
    For an $N$-body $d$-dimensional quantum system, there exists a set of states called the absolutely maximally entangled (AME) states~\cite{Helwig_2012} which find applications in various quantum information tasks such as teleportation, secret sharing, error-correction~\cite{Goyeneche_2015}, holography, and AdS/CFT correspondence~\cite{Pastawski_2015}. The distinguishing property of these states is that these are multipartite generalizations of the bipartite maximally entangled states. In other words, an AME state is maximally entangled over all bipartitions of size $\lfloor{\frac{N}{2}}\rfloor$. For example, an AME state $\ket{\Psi}_{1234} \in \mathcal{H}_1^{d_1}\otimes\mathcal{H}_2^{d_2}\otimes\mathcal{H}_3^{d_3}\otimes\mathcal{H}_4^{d_4}$ is maximally entangled with respect to all partitions $12|34$, $13|24$, and $14|23$, where $\{d_i\}$ are the local dimensions of the respective Hilbert spaces. An absolutely maximally entangled state $\ket{\Psi}\in \otimes_i^N \mathcal{H}_i^{d}$ of $N-$subsystems with local dimension-$d$ is represented as AME$(N,d)$.

    \begin{svgraybox}
        The explicit construction of an AME state of four qutrits, AME(4,3), can be written as
    \begin{align}
            % \ket{\Omega_{4,3}} = \dfrac{1}{3} \sum_{i,j = 0,1,2} \ket{i}\ket{j}\ket{i+j}\ket{i+2j}, 
            \centering
            \ket{\Omega_{4,3}} = \frac{1}{3} &(\ket{0000} + \ket{0112} + \ket{0221} 
            + \ket{1011}\nonumber \\  &+ \ket{1120} + \ket{1202} 
            + \ket{2022}+ \ket{2101}+ \ket{2210}).
    \end{align}
One can verify that all the reduced density matrices to two qutrits are $\hat{\rho} = \frac{1}{9}I_9$, with entropy $S = 2\text{log}3$ for every bipartition satisfying the condition of AME states. One of the \emph{open problems} with respect to AME states is the classification of the states in equivalence classes of entanglement~\cite{Rather_2023} (discussed later in Section~\ref{sec:entanglementresource}).
        
    \end{svgraybox}

    Since the states are maximally entangled over the bipartitions, the reduced density matrices are maximally mixed, which is analogous to the condition of obtaining maximally mixed states when taking the partial trace of any of the subsystems in the maximally entangled bipartite states.  It is known that AME states generally do not exist for every possible combination of $(N,d)$. There are certain combinations where no AME states exist,\footnote{An elaborative list of discovered AME states is available at ``\href{http://www.tp.nt.uni-siegen.de/+fhuber/ame.html}{http://www.tp.nt.uni-siegen.de/+fhuber/ame.html}''} and one of the main focus of the current research is to find all such configurations~\cite{Huber_2018}. One of the most recent breakthroughs in this direction was to show the existence of AME(4,6) by drawing a parallel with the age-old famous problem of 36 officers of Euler, by constructing quantum orthogonal Latin squares of this size~\cite{Rather_2022}.
\end{example}

The next step is to relax the condition of purity and account for the mixing of states (using the density matrix formalism as described in the previous section), which further generalizes the definition of separability and partial separability to mixed states.\\

\noindent\textbf{Separability of mixed states}.-- The notion of full separability for pure states from  Eq.~\eqref{Eq:pure_fullseparable} can be extended to mixed states, and a mixed separable state of a multipartite system is written as
    \begin{equation}\label{full-separability}
        \hat{\rho} = \sum_{i} p_i\hat{\rho}_i^{(1)}\otimes\hat{\rho}_i^{(2)}\otimes...\otimes\hat{\rho}_i^{(N)},
    \end{equation}
for the set of local density matrices $\hat{\rho}_i^{(j)}$ with probability distribution $p_i$~\cite{Werner1989, Dur_2000_sep}.

Similar to the bipartite case, all product states of the form $\hat{\rho}^{(1)}\otimes ... \otimes \hat{\rho}^{(N)}$ are separable, but the reverse is \emph{not} true, as we allow for a mixture of states.
Since there are many ways to create partitions in the mixed state multipartite scenario, this gives rise to the notion of partial separability where a state is not separable under any partition and yet not genuinely entangled. Any two states belong to the same \textit{separability class} if they are separable under the same set of partitions.\\

\noindent\textbf{Partial separability in mixed states}.-- The state $\hat{\rho}$ is separable under partitions $\{I_1, I_2,..., I_k\}$ such that $\cup_i^k I_i = I$ where $\{I_l\}$ are the disjoint subsets of indices, with $I = \{1,2,3,...,N\}$~\cite{Horodecki_2009}. The partially separable state has the form
\begin{equation}\label{partial_sep}
        \hat{\rho} = \sum_i p_i\hat{\rho}^{(I_1)}_i\otimes\hat{\rho}^{(I_2)}_i\otimes...\hat{\rho}^{(I_k)}_i,
\end{equation}
where any state $\hat{\rho}^{(I_j)}_i$ is defined on the tensor product of all elementary Hilbert spaces corresponding to indices belonging to set $I_j$.
Surprisingly, the conditions for full separability are stronger than separability with respect to all bipartitions -- there exist mixed states which are biseparable but still entangled~\cite{Bennett_1999_2}.\\

The concept of partial separability automatically gives rise to the notion of \textit{k-body} entanglement and entanglement depth which will be discussed in detail in the following Section~\ref{sec:4}.\\

Having defined the notion of entanglement in the multipartite setup and the complexity attached to it, the next step is to study the classification of states into various entanglement classes and class invariants. To do this, we consider the resource-theoretic framework of entanglement where entanglement is considered as a resource.

\subsection{Multipartite entanglement as a resource}
\label{sec:entanglementresource}
Entanglement as a resource has been studied extensively over the past few years~\cite{Chitambar_2019, ContrerasTejada_2019,Schwaiger_2015}, which has led to the development of a very elaborate and detailed resource theory of entanglement. Like any other resource theory, there are resource-free operations which are physical (or permissible) operations, and resource-free states which can be obtained from the action of resource-free operations. When considering entanglement as a resource:
\begin{itemize}
    \item \emph{Resource free operations}: entanglement non-increasing operations or physically realizable operations (specifically, local operations with classical communication-LOCC). The resource content of a state can be quantified using entanglement monotones (convex hulls of local entropy), i.e., non-increasing functions under resource-free operations.
    Usually resource free operations form convex sets, although extensions to non-convex sets have also been studied recently~\cite{Kuroiwa_2024,Salazar2024}.
    \item \emph{Resource free states}: separable states which can be generated from the resource-free operations. An entangled state is automatically a resourceful state.
\end{itemize}

The resource theory of entanglement provides a framework for better characterization and quantification of entanglement, and studies the manipulation and transformation of quantum states, and their utilization for specific quantum information tasks. Since entanglement is a purely non-local quantum phenomenon, any local operation \emph{cannot} increase the entanglement of the system. We start with the simplest class of resource-free operations which are the local unitary operations acting on each of the subsystems~\cite{Bennett_2000}.\\

\noindent\textbf{Local unitary (LU) operations}.-- 
Any two state vectors $\ket{\Psi}, \ket{\Phi}$ are considered to be equivalently LU-entangled if they differ by a local unitary basis change:
\begin{equation}
    \ket{\Psi}\sim_{LU}\ket{\Phi} \equiv \ket{\Psi} = (U_1 \otimes U_2 \otimes ... \otimes U_N)\ket{\Phi},
\label{eq:LU}
\end{equation}
where $U_i$ are the local unitary operations acting on the respective parties. Having defined the LU-equivalence relation, the next step is to classify the set of entangled states into LU-inequivalent classes.\\

For the bipartite case, this classification into entangled and non-entangled states is much simpler due to the notion of Schmidt coefficients as introduced in the previous section. Here, we will revisit the topic from the perspective of LU-equivalence:
A bipartite pure state can be expressed in the form
    \begin{equation}
        \ket{\Psi} = \sum_{i=1}\sqrt{p_i} \ket{\psi_i}\otimes\ket{\phi_i},
    \end{equation}
and the action of local unitaries $U_1$ and $U_2$ on the respective subsystems results in the basis change of the subsystems shown as
    \begin{equation}
        (U_1 \otimes U_2)\ket{\Psi} = \sum_{i=1}\sqrt{p_i}(U_1\ket{\psi_i})\otimes (U_2\ket{\phi_i}).
    \end{equation}
The Schmidt coefficients\footnote{There is a physical interpretation of Schmidt coefficients in the bipartite case as they are the set of eigenvalues of each of the reduced density matrices of the bipartite system.} $\{p_i\}$ are invariant under the action of local unitaries. The LU-inequivalent classes are completely described by these Schmidt coefficients in the bipartite case.

However, for the multipartite case, the characterization becomes more challenging.\footnote{There is no singular value decomposition for tensors of higher degree than two.} This is evident from counting the number of parameters necessary to describe a vector of $N$-qubits in the quotient space with respect to the given inequivalent relation.

\begin{example}{LU-parameter counting in $N$-qubits}
    It takes $2^{N+1} - 2$ real parameters to specify a normalized quantum state in $\mathcal{H} = (\mathcal{C}^2)^{\otimes N}$, whereas the group of local unitary transformations $\text{SU}(2) \times ... \times \text{SU}(2)$ has $3N$ real parameters. This means that even in the case of $N-$qubits, one needs at least $2^{N+1}-3N-2$ real numbers to parameterize the sets of LU-inequivalent pure quantum states~\cite{Linden_1998, Carteret_2000, Walter_2016}. Unlike the Schmidt coefficients from the bipartite case, most of these parameters do not have a physical interpretation.
\end{example}

Another key difference in the multipartite scenario is that only very rarely pure multipartite states admit the generalized Schmidt decomposition~\cite{Peres_1995, Thapliyal_1999} unlike the bipartite case.

% An example of states that can be written in the generalized Schmidt decomposition are the GHZ-states defined as \cite{Horodecki_2009}
% \begin{equation}
% \ket{\text{GHZ}}_d^{(N)} = \frac{1}{\sqrt{d}}\sum_{i=0}^{d-1}\ket{i}^{\otimes N},
% \end{equation}
% for the case of $N-$partite systems of local dimension $d$. \\

In contrast to local unitary operations, \emph{global} unitary ones are able to create entanglement. 
Such features were quantified as \emph{entangling power}, which measures the (average or maximal) entanglement created on separable pure states, studied for both bipartite systems~\cite{Zanardi_2000,Eisert_2021}, as well as in the multipartite case~\cite{Linowski_2020}  
As it turns out, its counterpart -- the disentangling power, which measures how the entanglement degrades upon the action of the operation is closely related in certain cases~\cite{Clarisse_2007}, but not in general~\cite{Linden_2009}.

Now we shall extend the realm of unitary operations.
Apart from LU, there exists a wider class of operations that cannot generate entanglement from separable states. 
It will give rise to a coarser notion of ``equivalent entanglement''.\\
% Since entanglement is a purely non-local \emph{quantum} phenomena, one cannot generate entanglement by the action of local unitary operations from separable states. We can obtain a coarser degree of ``equivalent entanglement" by considering larger set of operations which still satisfy this property.\\

\noindent\textbf{Local operations and classical communication (LOCC)}.-- 
A more general form of the LU operations which can still be completely defined classically are the local operations with classical communication (LOCC). LOCC allows for an exchange of classical information of the local measurement outcomes of the respective parties, hence, the name. Thus, the choice of the local operations of individual parties can be affected by the information of measurement outcomes by any other party. The best way to imagine this is with the help of the ``distant laboratories model'' \cite{Chitambar_2014}. There are $N$-particles each in their laboratory. The particles may or may not be entangled in the first place. Each laboratory is capable of performing local measurements on its own particle, and conveying the information regarding the outcome without exchanging the quantum systems among themselves. This whole process is called one round of LOCC, where all parties perform measurements based on the previously circulated information, and further pass on the information regarding the current measurement outcomes to the rest of the parties. The LOCC is difficult to characterize, but there exists a set of general operations called separable operations SEP, such that $\text{LOCC} \subset \text{SEP}$, which can be represented as a channel $\Lambda$ acting as
\begin{equation}
    \Lambda (\hat{\rho}) = \sum_a \left(\otimes_{i=1}^N K_{a,i}\right) \hat{\rho}\left(\otimes_{i=1}^N K_{a,i}\right)^{\dagger},
\end{equation}
where $K_{a,i}$ are the local Kraus operators acting on the $i$-th subsystem respectively~\cite{Bennett_1999, Donald_2002}. This is a direct generalization of LU operations, where we considered the unitary matrices of Eq.~\eqref{eq:LU}, which are composed of only one Kraus operator, acting on the subsystem $i$, given by $U_{i} = K_{a=1,i}$.

\begin{svgraybox}
    LOCC are considered the resource-free operations in the resource theory of entanglement~\cite{Bennett_1996_v3, Bennett_1996_v2, Vedral_1997} since any separable state $\ket{\Psi} \in \text{SEP}$ can be obtained from any other state in the Hilbert space by the action of LOCC. On the other hand, it is impossible to generate any entangled state from a separable state through LOCC alone~\cite{Bennett_1999}.
\end{svgraybox}

A typical LOCC protocol has many rounds,\footnote{Some of the maps require infinite rounds of LOCC for implementation~\cite{Chitambar_2011, Chitambar_2014}.} where each laboratory performs a (POVM) measurement on their particle. The post-measurement state is stored and the classical outcome is broadcasted to the other laboratories. In the next round, another party performs measurement on their particle based on the results of the previous measurement outcomes, and so on. Any two states are LOCC-\emph{equivalent} if they can be interconverted by this kind of protocol. For mathematical simplicity, there are several variants of this definition based on the allowed number of rounds, for example, two states are LOCC$_r$-equivalent if they can be interconverted using LOCC protocol with no more than $r$ rounds. They are $\overline{\text{LOCC}}$-equivalent if starting from any of them, one can approximate to the other one with arbitrary precision if the number of rounds $r\rightarrow\infty$.

Although having a physical interpretation, there is no simple mathematical description of LOCC-equivalence in the multipartite scenario so far. There is no known algorithm that decides whether two vectors are $\overline{\text{LOCC}}$-equivalent, even after an exponential runtime in the total dimension \cite{Chitambar_2014, Walter_2016}. However, in the case of bipartite entanglement, there exists a well-defined mathematical framework for studying the LOCC-equivalence classes, using the majorization condition of Schmidt coefficients (Nielsen's theorem)~\cite{Nielsen_1999}:

\begin{overview}{Nielsen's theorem}
Let $\ket{\Psi}, \ket{\Phi}$ be two states in $\mathbb{C}^d\otimes\mathbb{C}^d$ with respective Schmidt coefficients $\boldsymbol{\lambda} =\{\lambda_i \}, \boldsymbol{\mu} = \{\mu_i \}$ in increasing order. Then, there exists a LOCC protocol converting $\ket{\Psi}$ exactly to $\ket{\Phi}$ if and only if $\boldsymbol{\lambda}$ is majorized by $\boldsymbol{\mu}$, i.e., $\forall k \in \{1,2,..,d \}$,

\begin{equation}\label{majorize}
\sum_{i=1}^k\lambda_i\leq\sum_{i=1}^k\mu_i.\\
\end{equation}

There exist some generalizations of Nielsen's theorem to the multipartite scenario as shown in Ref.~\cite{Gour_2011}.
\end{overview}

% In the multipartite setting, there exists a set of states that are \emph{maximally useful} in the sense that any state outside the set can be obtained via LOCC from one of the states inside the set, and no state in the set can be obtained from any other states in the set via LOCC analogous to the properties of maximally entangled set in the bipartite case. \aks{"In the multipartite setting, there exist maximally entangled states, such that each has an equivalance class associated to it."

% "Any state from the equivalance class can be generated from the corresponding maximally entangled state via LOCC".\\

We can further have a stricter criterion by allowing for a level of randomness in local transformations. This classification scheme goes beyond the deterministic nature of LOCC, offering a more detailed understanding of the entanglement classification in multipartite systems.\\

\noindent\textbf{\emph{Stochastic}-LOCC (SLOCC)}.--
Any two states are SLOCC equivalent if they can be converted into each other by LOCC with some finite probability. Similar to LOCC, the SLOCC protocol consists of several rounds with each party performing operations based on measurement outcomes by other parties. The whole process can be imagined as a tree, and every measurement results in a new branch of the tree. If at least any one of the branches leads to the target state $\ket{\Phi}$ starting from $\ket{\Psi}$, the two states are SLOCC-equivalent.\\

\begin{itemize}
    \item A conversion from $\ket{\Psi}\xrightarrow{\text{SLOCC}}\ket{\Phi}$ is possible under SLOCC if there exists a set of operators $\{A_i\}$ (compared to Eq.~\eqref{eq:LU}) such that
\begin{equation}
    \ket{\Psi}\sim_{\text{SLOCC}}\ket{\Phi} \equiv \ket{\Psi} = (A_1 \otimes A_2 \otimes ... \otimes A_N)\ket{\Phi},
    \label{eq:SLOCC}
\end{equation}
    \item In particular, the two states $\ket{\Psi} \xleftrightarrow{\text{SLOCC}}\ket{\Phi}$ are SLOCC-equivalent iff the matrices $\{A_i\}$ are invertible (or determinant $\text{det}\ A_i \neq 0$)~\cite{Dur_2000, Verstraete_2002}. The matrices $\{A_i\}$ for which $\text{det}\ A_i = 1$ which satisfy this property form the \emph{special linear} group SL.
\end{itemize}
Similar to the case of $N-$qubits with LU-equivalence, one obtains a lower bound on the number of parameters required to define the SLOCC-equivalence classes given by $2^{N+1} - 6N - 2$ by substituting the $\text{SU}(2)$ group with $\text{SL}(\mathbb{C}^{2})$ group.\\

Now that we have defined the operations and transformations that are allowed within the resource theory framework of entanglement, we consider the example of three qubits to study the classification of states and the corresponding invariants for each of the classes.\\

\begin{example}{Entanglement classification of 3 qubits}
The simplest model to extend our discussion on multipartite entanglement is a tripartite system of three qubits. Based on the equivalence relations shown above with respect to different types of operations, quantum states of 3-qubits can be classified into six SLOCC-inequivalent entanglement classes~\cite{Dur_2000}.

Before advancing the discussion further toward the classification of 3-qubit entangled states, let us define the notion of class invariants.

\begin{definition}
    SLOCC-invariants are referred to as the functions of the states that do not change under the action of SLOCC. These invariants play a crucial role in characterizing and distinguishing between different types of entangled states within the SLOCC framework. Tensor rank and hyperdeterminant are the two well-known and studied SLOCC-invariants that we will later use to classify the genuinely tripartite entangled states. 
\end{definition}

We study the classification of states based on these invariants. First, we restrict our discussion to 3-qubit pure states and later generalize it to mixed states to obtain a hierarchical structure of these sets informally known as the \emph{onion ring structure}~\cite{Dur_2000, Acin_2001, Acin_2001_mixed}.

\begin{itemize}
    \item The first class of states that exist are the set of separable states (or free states) written as product states of the form $\ket{\Psi}  = \ket{\psi_1}\otimes\ket{\psi_2}\otimes\ket{\psi_3}$. In this particular case of separable states, $r(\rho_1) = r(\rho_2) = r(\rho_3) = 1$, where $r(\rho_i)$ is the local rank of the reduced density matrix of $i$-th subsystem. The rank is invariant under the action of invertible SLOCC~\cite{Dur_2000, Dur_2000_sep, Acin_2001, Walter_2016}.\\

    \item Next, there are three classes of bipartite states based on the bipartitions $1|23, 2|13,$ and $3|12$. For each bipartition $i|jk$, the set of pure product states forms an SLOCC class,\footnote{This follows from the fact that any two entangled pure states of two parties are equivalent under SLOCC.} and are of the form $\ket{\psi_i}\otimes\ket{\Phi_{j,k}}$, where $\ket{\Phi_{j,k}}$ is a non-product state of subsystems-$j,k$, where $r(\rho_i) = 1$, and $r(\rho_{j}) =r(\rho_{k}) = 2$. Thus, giving rise to three separate SLOCC-inequivalent classes of entanglement~\cite{Dur_2000}.\\

    \item Finally, the remaining states are 
    genuinely 3-qubit entangled -- of maximal local rank with respect to any single-qubit $r(\rho_1) = r(\rho_2) = r(\rho_3) = 2$.
    Nonetheless, among these states, one can distinguish two inequivalent classes~\cite{Dur_2000, Acin_2001}. These two classes of states are the W and the GHZ states, which are discussed below in detail, and shown to be SLOCC-inequivalent using invariants such as hyperdeterminant and tensor rank.\\
\end{itemize}

\begin{example}{Using rank as an invariant in bipartite systems}
    In the bipartite scenario, two state vectors $\ket{\Psi}$ and $\ket{\Phi}$ can be written in the product basis using coefficient matrices $T_{i,j}$ and $T'_{i,j}$ as follows:
    \begin{equation}
        \ket{\Psi} = \sum_{i,j}^d T_{i,j} \ket{i}\otimes\ket{j}, \ \ \ \ \ \ \  
        \ket{\Phi} = \sum_{i,j=1}^d T'_{i,j}\ket{i}\otimes\ket{j}.
    \end{equation}

    Let us assume $\ket{\Psi}$ and $\ket{\Phi}$ are SLOCC-equivalent.
    Then, there exist invertible operators $A_1$ and $A_2$ such that $A_1\otimes A_2 \ket{\Psi} = \ket{\Phi}$ as shown in Eq.~\eqref{eq:SLOCC}, and ranks of the coefficients are equal $r(T') = r(T)$~\cite{Walter_2016}. Hence, the rank of the coefficient matrix is invariant under invertible SLOCC operations.

    \begin{svgraybox}
    A pure state of two qu$d$its is entangled with maximal Schmidt rank iff $\text{det}\ T \neq 0$.
\end{svgraybox}
\end{example}

\noindent\textbf{Hyperdeterminant}.-- Hyperdeterminant is the generalization of the determinant defined for the matrices, to higher order tensors~\cite{Gelfand_1994}. In 2003, Miyake~\cite{Miyake_2003} showed that the entanglement measures such as concurrence and tangle are special forms of hyperdeterminant, and, as such, are invariant under LU operations. Concurrence in the bipartite case is the determinant of the coefficient matrix of a state $C(\Psi) = 2|\text{det}\ T|$, whereas tangle is the hyperdeterminant of second order, which is given by the $\tau_3 (\ket{\Psi}) = |\text{det}\ \Psi|$~\cite{Coffman_2000, Osterloh_2006}. 
The strict positivity of the hyperdeterminant is an invariant under the local general linear group $\text{GL}^{\times N}$~\cite{Miyake_2003}. The hyperdeterminant is an example of LU-invariant which can be expressed as a polynomial function of the state's coefficients~\cite{Makhlin2002}.

It has been shown that the $\text{det}(\text{W})=0\neq \text{det}(\text{GHZ})$, thus giving rise to two different SLOCC-equivalent classes within 3-qubit genuinely tripartite entangled states. Therefore, from the perspective of \textit{3-tangle} (sum of tangles with respect to all bipartitions), the GHZ-states are more entangled than the W-states, thus detecting so-called \emph{monogamy} of entanglement~\cite{Coffman_2000}.\\

\noindent\textbf{Tensor rank}.-- We can also show the distinction between the GHZ and W SLOCC-inequivalent classes of pure states using tensor rank as an invariant. Let us decompose any pure state vector using the minimal decomposition
    \begin{equation}
        \ket{\Psi} = \sum_{i=1}^{R_{\text{min}}} c_i \ket{\psi_i^{(1)}}\otimes ... \otimes \ket{\psi_i^{(N)}},
    \end{equation}
where $\left\{\ket{\psi_i^{(j)}}\right\}_{i=1}^{R_{\text{min}}}$ may or may not be orthogonal unlike the Schmidt decomposition (where all the vectors need to be pair-wise orthogonal to each other). The number of terms $R_{\text{min}}$ defines the tensor rank\footnote{The logarithm of tensor rank is also called the \textit{Schmidt measure}.} of the state $\ket{\Psi}$. 
Unlike the Schmidt rank for bipartite systems, for the multipartite case, the tensor rank is not easy to compute.
It is invariant under SLOCC operations, and it distinguishes the two genuine tripartite entangled SLOCC-inequivalent classes of W and GHZ states which can be shown below:

The state vectors representing GHZ~\cite{Greenberger_1989} and W~\cite{Dur_2000} states respectively are
\begin{equation}
    \ket{\text{GHZ}} = \frac{1}{\sqrt{2}}\left( \ket{000} + \ket{111}\right), \ \ \ \  
    \ket{\text{W}} = \frac{1}{\sqrt{3}}\left( \ket{001}+\ket{010}+\ket{100}\right).
\end{equation}
Based on the definition of tensor rank, we can see that it is not possible to express $\ket{\text{W}}$ state using only two product terms unlike $\ket{\text{GHZ}}$ state, i.e., $R_{\text{min}}(\ket{\text{GHZ}}) = 2$ and $R_{\text{min}}(\ket{\text{W}}) = 3$, hence they cannot be interconverted using SLOCC alone. However, pure states from the W class can be approximated to an arbitrary precision by states in the GHZ class~\cite{Acin_2001, Vrana_2015}, while the converse is not true.\\
%, indicating that GHZ states are more entangled than the W states in some sense.\\

Based on the discussion above, we can represent the different SLOCC-inequivalent classes in Table~\ref{tab:sloccthreequbits} and the corresponding invariants used to distinguish the respective classes. Here, we show the local entropy for the reduced subsystems as described in Def.~\ref{def:entropy}, local rank, and the 3-tangle $\tau$ as the class invariants for each of the six SLOCC classes.
This set of invariants is complete -- it allows us to fully distinguish between different LOCC classes in the case of pure states.\\

% \begin{table}[h]
% \label{tab:sloccthreequbits}
%  \begin{center}
% \begin{tabular}{|p{1cm}|p{0.7cm}|p{0.7cm}|p{0.7cm}|p{0.7cm}|}
%  \hline

% Class & $S_1$ & $S_2$ & $S_3$ & $\tau$ \\
%  \hline
% SEP & 0 & 0 & 0 & 0 \\ 
% $1|23$ & 0 & $>0$ & $>0$ & 0 \\ 
% $2|13$ & $>0$ & 0 & $>0$ & 0 \\ 
% $3|12$ & $>0$ & $>0$ & 0 & 0 \\
% W & $>0$ & $>0$ & $>0$ & 0 \\
% GHZ & $>0$ & $>0$ & $>0$ & $>0$ \\
%  \hline

% \end{tabular}
% \end{center}
% \caption{Values of local entropies $S_1, S_2, S_3$ of the reduced subsystems, and the 3-tangle $\tau$ for different SLOCC-inequivalent entanglement classes}

% \end{table}

\begin{table}[h]
\label{tab:sloccthreequbits}
 \begin{center}
\begin{tabular}{|p{1cm}|p{0.7cm}|p{0.7cm}|p{0.7cm}|p{0.7cm}|p{0.7cm}|p{0.7cm}|p{0.7cm}|}
 \hline

Class & $S_1$ & $S_2$ & $S_3$ & $r_1$ & $r_2$ & $r_3$ & $\tau$\\
 \hline
SEP & 0 & 0 & 0 & 1 & 1 & 1 & 0 \\ 
$1|23$ & 0 & $>0$ & $>0$ & 1 & 2 & 2  & 0 \\ 
$2|13$ & $>0$ & 0 & $>0$ & 2 & 1 & 2 & 0 \\ 
$3|12$ & $>0$ & $>0$ & 0 & 2 & 2 & 1 & 0 \\
W & $>0$ & $>0$ & $>0$ & 2 & 2 & 2 & 0 \\
GHZ & $>0$ & $>0$ & $>0$ &  2 & 2 & 2 & $>0$ \\
 \hline

\end{tabular}
\end{center}
\caption{Values of local entropies $\{S_1, S_2, S_3\}$ and the local ranks $\{r_1, r_2, r_3\}$ of the reduced subsystems, and the 3-tangle $\tau$ for different SLOCC-inequivalent entanglement classes.}

\end{table}

\noindent\textbf{Mixed states}.-- The situation is more involved when we allow for a mixture of pure states -- now not all states from a given class will be SLOCC interconvertible, but they can be created using a mixing of states from an appropriate class of pure state entanglement, up to an arbitrary precision. %mixing of the states of different entanglement classes as it is even more difficult to classify an arbitrary mixed state into one of the classes. 
The hierarchical structure of entanglement classes in the mixed case scenario for 3 qubits~\cite{Bennett_1999_2, Dur_2000, Dur_2000_sep,  Plenio_2001, Acin_2001, Acin_2001_mixed, Eisert2001} has been shown to be
\begin{equation}\label{eq:inclusion_3_qubit_mixed_classes}
    S \!\subset\! B \!\subset\! W \!\subset\! GHZ,
\end{equation} 
where in this context:
\begin{itemize}
    \item \textit{S} is the class of all convex sum of projectors onto pure separable vectors,
    \item \textit{B} is the closure of the set defined as the convex sum of projectors onto pure biseparable entangled vectors (from all bipartitions: $1|23, 2|13, 3|12$),
    \item \textit{W} is the closure of the set defined as the convex sum of projectors onto pure W states (in contrast to the pure case, $W$ class is not of measure zero.),
    \item \textit{GHZ} is the set of all possible physical states, i.e., the closure of the set containing convex sums of projectors onto pure GHZ states.
\end{itemize}

The reason for the introduction of the closure is that all states in $S$ have a biseparable state\footnote{More precisely, a state formed as a convex sum of projectors onto biseparable states. The same remark applies to other neighborhoods as well.} in the neighborhood, all states in $B$ have a W state in the neighborhood, and all states in $W$ have a GHZ state in the neighborhood~\cite{Dur_2000}. 
However, there are states that belong to only one of those classes, e.g., W states that cannot be written a convex combination of projectors onto GHZ states -- the simplest example are all pure states.
We emphasize this distinction in Figure~\ref{fig:entclass3}, highlighting the internal structure of the biseparable class to provide more geometrical intuition.

\begin{figure}[h!]
    \centering
    \includegraphics[width = .75\linewidth]{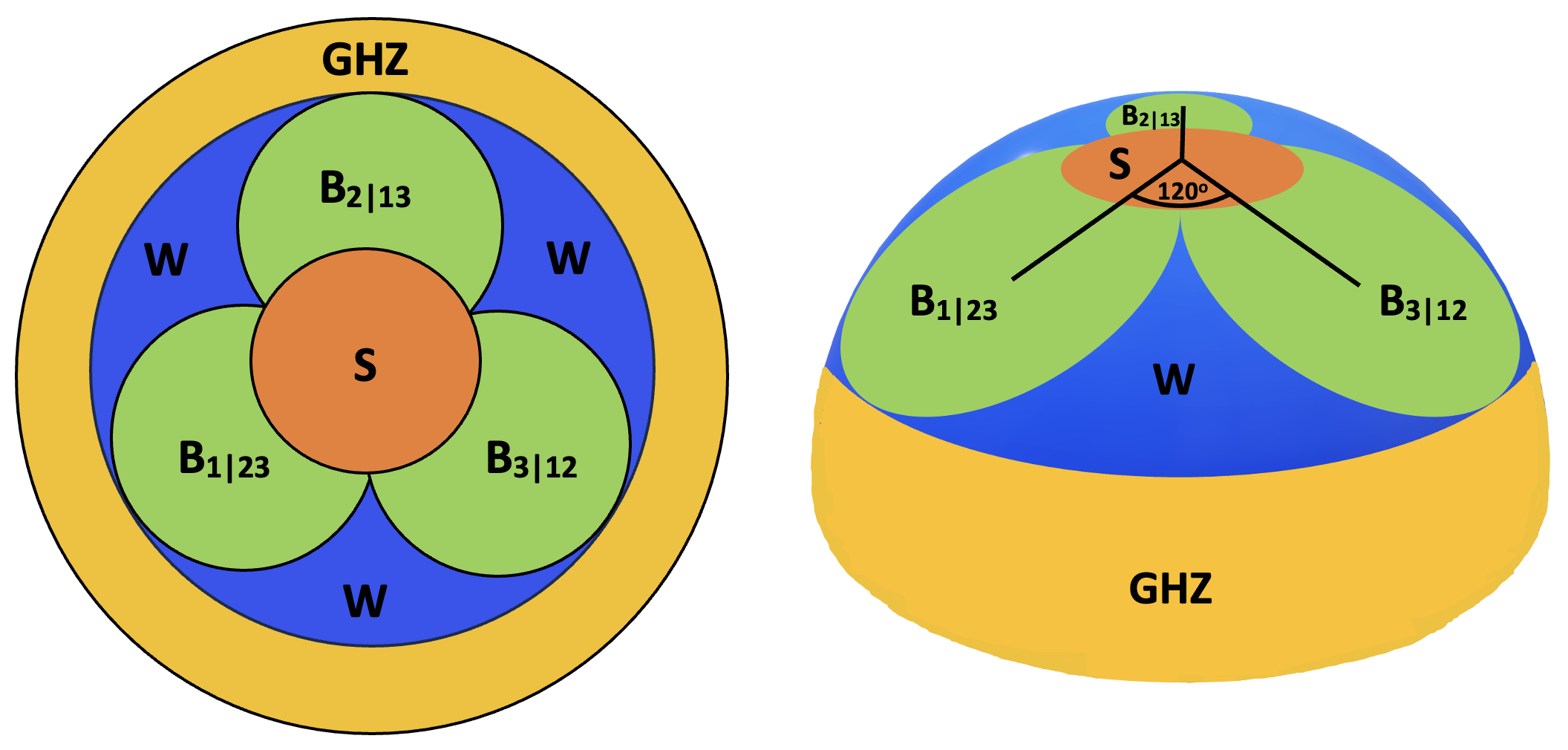}
    \caption{The four possible entanglement classes based on the SLOCC operations in the case of tripartite systems. Such sketches were done many times before~\cite{Acin_2001_mixed, Guhne_2009}, but we believe that a 3-dimensional structure is needed to fully understand the intricacies of the set -- W class is smaller than GHZ, but there are W states that cannot be written as a convex combination of GHZ states alone~\cite{Acin_2001_mixed}.
    Therefore, the inclusion (\ref{eq:inclusion_3_qubit_mixed_classes}) requires the closure of the sets -- any W state can be \emph{approximated} by GHZ states up to arbitrary precision, same as biseparable states can be approximated via W states and separable ones via biseparable. 
    Note that the separable set $S$ is a proper subset of the intersection of $B_{1|23} \cap B_{2|13} \cap B_{3|12}$; in other words, there exist states that are separable with respect to every bipartition yet entangled~\cite{Bennett_1999_2}.
    We express our gratitude to Karol \.Zyczkowski for his valuable remarks regarding the figure.}
    \label{fig:entclass3}
\end{figure}

Characterization of those classes can be performed using entanglement witnesses~\cite{Acin_2001_mixed} via best separability approximation~\cite{Lewenstein_1998}, utilizing the Schmidt number~\cite{Eisert2001, Sanpera_2001}, or via Bell correlations~\cite{Plodzien2024_Ent_class}. 
Alternatively, multipartite entanglement classification for pure states involves studying the local purities of the reduced states. 
Such methods produce similar pictures to Figure~\ref{fig:entclass3}, yielding an object known as the entanglement polytope or the Kirwan polytope~\cite{Walter_2013,Sawicki_2013,Sawicki_2012,Sawicki_2014,Maciazek_2018,Enriquez_2018}. 
However, for a higher number of parties, it is a rudimentary division, not allowing to highlight the intricacies of different entanglement classes. 
% This classification leads to the hierarchical structure of the classes as shown in the Figure~\ref{fig:entclass3}.

\end{example}

Increasing the number of parties, the natural question is the entanglement classification for four qubits (four parties). The difficulty to characterize entanglement increases as the number of parties grows -- already for four qubits there are infinitely many SLOCC classes~\cite{Verstraete_2002}.

Since there does not exist a generalized theory of multipartite entanglement, it is more feasible to focus on concrete families of states of physical relevance. 
An example of them are matrix product states, which efficiently approximate ground states of one-dimensional quantum many-body systems and will be studied in the next section. 
Another approach is to characterize many-body entanglement from a given set of accessible few-body expectation values.
This so-called \textit{data-driven} approach will also be touched upon in the following section.

\section{Use and detection of many-body quantum entanglement}
\label{sec:4}
\textit{Certainly, there is more to this world than just a few particles}: systems composed of many interacting constituents or $\textit{bodies}$ are ubiquitous in Nature. They can be found in the whole range of the energy scale, from elementary particles forming atoms, atoms bindings to form molecules, many atoms interacting in solid state materials or gases, planetary systems, and beyond. Such systems are intrinsically challenging to describe. Moreover, they feature collective phenomena in the form of emergent dynamics which are hard to predict from the properties of the individual constituents.  

One may think that interactions will eventually decohere the system, washing out any possible quantum contribution to thermal noise. This is the norm when the classical description is sufficient as the microscopic degrees of freedom can be integrated out. However, some emergent phenomena require a quantum description, with entanglement playing a central role. Such systems include Bose-Einstein condensation at finite temperature \cite{Anderson_1995} and exotic phases of matter at zero temperature driven by quantum fluctuations \cite{Zeng_2019,Sachdev_2023}. Remarkably, these experiments show that quantum physics can be manifested macroscopically and witnessed ``with the naked eye'', beyond its traditional scale of applicability. \\

\noindent\textbf{The many-body regime}.-- We consider a closed system formed by $N$ elementary constituents (particles). In the many-body regime, the collective dynamics dominate, masking any detailed local degrees of freedom from being \textit{relevant} in the observers' macroscopic scale. A necessary condition for the collective behaviors to take over is that $N$ must be sufficiently large\footnote{Such number will generally depend on the nature of the interactions and can be as high as the Avogadro constant, $N_A \sim  10^{24}$. } such that the observed quantities (with a given scaling) become independent of $N$. For the purpose of this \textit{section}, though, the main message is that for most practical applications, such $N$ is too high for a generic (entangled) many-body state to be stored and processed in an ordinary computer:

\begin{overview}{Curse of dimensionality}
The state of $N$ interacting particles with $d$ internal degrees of freedom (e.g. spin) is written in the computational basis as,  
\begin{equation}
    \label{eq: MB_def}
    \ket{\Psi} =\sum_{\{s\in [d]\}} \Psi_{s_1s_2..s_N}\ket{s_1,s_2,...,s_N}\in [\mathbb{C}^d]^{\otimes N} \ .
\end{equation}
Notice how $\order{d^N}$ parameters are needed to just specify the state. This number quickly becomes unfeasible to store in a conventional computer.    
\end{overview}

\noindent However, not everything is lost, as:

\begin{svgraybox}
The hopes for scalable descriptions:
\begin{enumerate}
\item While generic states in the Hilbert space are highly entangled, those physically relevant are typically weakly entangled.
\item Even if the global state is inaccessible, partial tomography can be performed and it might be sufficient to witness its resource content (e.g. in the form of many-body entanglement or stronger quantum correlations). 
\end{enumerate}
\end{svgraybox}

\noindent In the remainder of the \textit{section}, we will formalize and elaborate on these two highly non-trivial aspects. We will show as well how despite the complexity of the scenario, entanglement can still be characterized and certified to be exploited in current applications. \\

\begin{figure}[h]
\sidecaption
% Use the relevant command for your figure-insertion program
% to insert the figure file.
% For example, with the graphicx style use
\includegraphics[width = 0.4\textwidth]{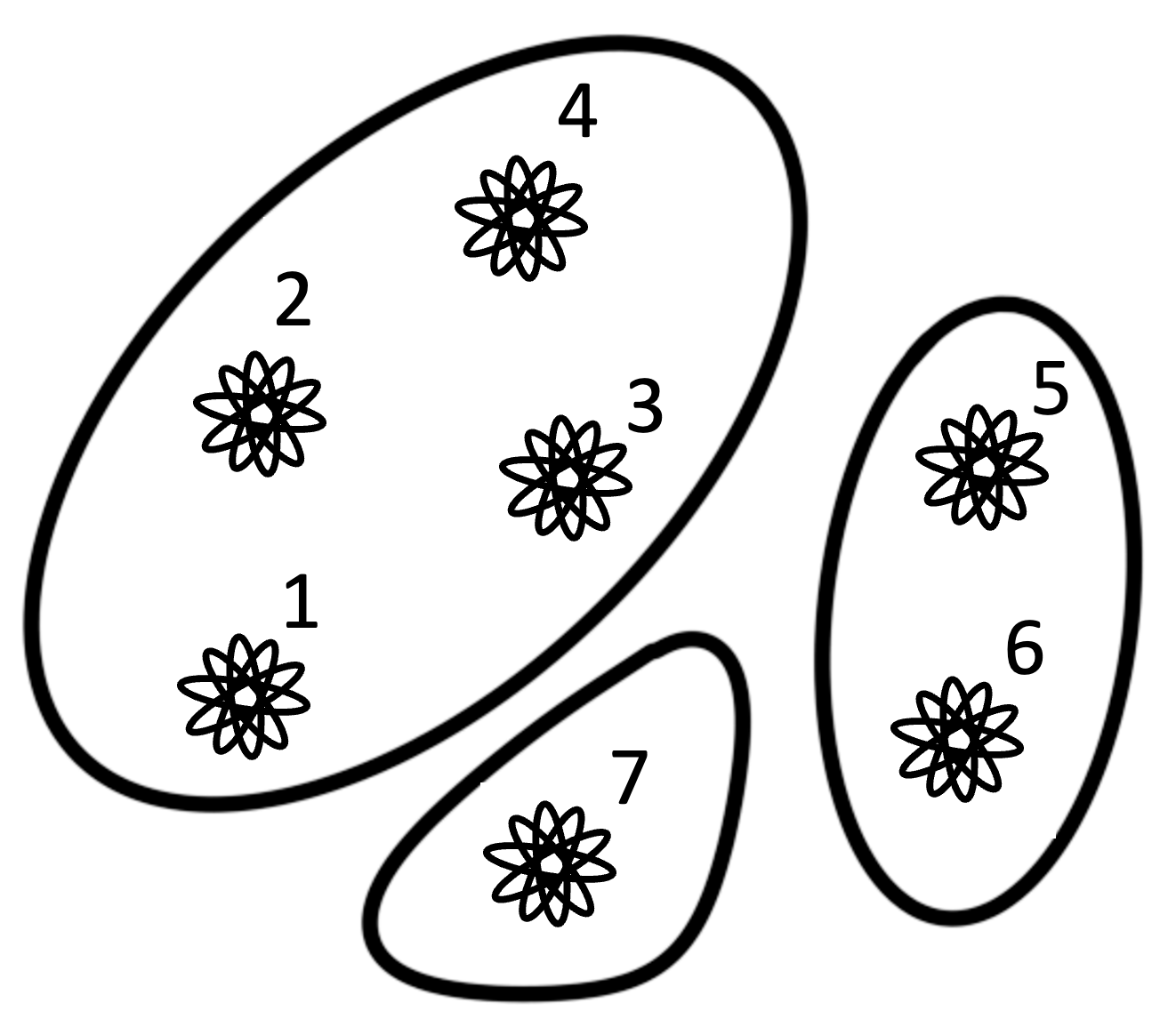}
%
% If no graphics program available, insert a blank space i.e. use
%\picplace{5cm}{2cm} % Give the correct figure height and width in cm
%
\caption{Collection of $N = 7$ distinguishable atoms and the partition $\{\{1,2,3,4 \},\{5,6\},\{7 \} \}$. Atoms in the same partition add to a global GME state, i.e. they cannot be further separated without degradation (losing part of the state's description). The entanglement depth refers to the minimal number of GME parties needed to describe the state. In this case: $K = 4$.}
\label{fig:depth}       % Give a unique label
\end{figure}

\noindent\textbf{Entanglement depth}.-- In the previous \textit{section}, we have studied entanglement with respect to different partitions or \textit{parties}. Here, as a complementary view, we focus on the number of subsystems or \textit{bodies}. There is a distinction between \textit{bodies} and \textit{parties} as defined here. In particular, many-body states can be studied with respect to different partitions. As a reminder, we define $\mathcal{P}_K$ as a partition of $[N]:=\{1,2,..,N\}$ in non-empty pairwise disjoint subsets of at most $K$ elements (see Figure~\ref{fig:depth}). A global vector $\ket{\Psi}\in \mathcal{H}^{\otimes N}$ may factorize in a given partition $\mathcal{P}_K$ as
\begin{equation}
    \ket{\Psi_K} = \bigotimes_{p\in\mathcal{P}_K}\ket{\psi_p}.
\end{equation}
If such decomposition exists, we say that $\ket{\Psi_K}$ is $K$\textit{-producible with respect to the partition} $\mathcal{P}_K$. Such a definition is extended to mixed states by the usual convex hull construction. We call a state $\hat{\rho}\in \mathcal{B}(\mathcal{H}^{\otimes N})$ $K$\textit{-producible} if it can be expressed as a convex combination $K$-producible states (possibly with respect to different partitions), i.e.,   
\begin{equation}
    \label{eq:Kprod_mixed}
    \hat{\rho}_{K} = \sum_{\Psi}p_{\Psi}\ketbra{\Psi_K},
\end{equation}
with $\{p_\Psi\}_{\Psi}$ a probability distribution.\footnote{If $\mathcal{H}$ is finite-dimensional, a sum is sufficient to represent classical mixing, although it has generally infinitely many terms. For continuous variables, if $\mathcal{H}$ is infinite-dimensional, the sum must be replaced by an integral.}  

Note that if $K=1$, then $\mathcal{P}_1 = [N]$ and $\hat{\rho}$ is \textit{fully separable} (SEP), that is, not entangled. In general, entanglement among $K$ parties is necessary to generate a $K$-producible state. From a resource point of view, generating multipartite entanglement is not for free. Therefore, one may be interested in the minimal $K$ for which a decomposition as per Eq. \eqref{eq:Kprod_mixed} exists. Such value is the so-called $\textit{entanglement depth}$ and quantifies the minimal number of genuine multipartite entangled parties required to describe the state \cite{Guhne_2005, Lucke_2014}. In particular, if $K=N$, then we say that the state is genuinely-multipartite entangled (GME). It is clear from the construction that the set of states with entanglement depth $K$, $\mathcal{R}_K$, is convex and $\mathcal{R}_k\subset\mathcal{R}_{k+1}$. Moreover, under local operations, the entanglement depth can only be reduced. which illustrates the many levels of depth in which a many-body state can be entangled, from fully separable to genuinely $N$-body entangled. 
As compared with the previous \textit{sections}, this classification is coarse: two states with the same entanglement depth may be locally inequivalent. However, such operations can not increase the value of $K$, which makes it a valid entanglement measure.

\subsection{Useful (and useless) many-body entanglement}

As explained in previous \textit{sections}, quantum entanglement is a resource. However, most of the quantum states are useless, in the sense that their power can not be exploited. Here, we will focus on some features of two classes of many-body states which have stood out for their physical relevance and usefulness in current applications: matrix product states and symmetric states.  

\subsubsection{Matrix product states}

\textbf{Many-body states are tensors}.-- We interpret the coefficients of the many-body state Eq. \eqref{eq: MB_def} as an  $N$-tensor $\Psi : [d]\times [d]\times ...\times [d]\rightarrow\mathbb{C}$ such that $( s_1,s_2,..,s_N)\mapsto\Psi_{s_1s_2..s_N}$. Notice how for separable $\Psi$, the tensor factorizes (compare with Eq.~\eqref{Eq:pure_fullseparable}),
\begin{equation}
\label{eq:MB_sep}
\Psi_{s_1s_2...s_N}=\psi_{s_1}\psi_{s_2}...\psi_{s_N} \ ,
\end{equation}
and only $\order{dN}$ coefficients are needed to be specified. Such a case is realized in noninteracting systems. As soon as interaction is switched on, entanglement is generated and the number of necessary parameters grows exponentially $\order{d^N}$. However, the interaction mechanisms realized in Nature are very particular. Which entanglement patterns are produced? Can they be efficiently handled? \\

The interactions are encoded in a Hamiltonian $\hat{H}$. Typically, its physical realizations possess the following basic properties \cite{Eisert_2010}: 

\begin{enumerate}
    \item \textbf{Locality}:
    Fundamental interactions happen locally,\footnote{Notice how e.g. the Coulomb law of electrostatics is nonlocal. However, it is an effective theory. The fundamental interaction consists in giving and receiving virtual photons, as a local process.} that is, within compact neighborhoods $\{ X\}$, with support independent from $N$, of a topological space $L$. Such geometry structures the Hamiltonian as: 
\begin{equation}
   \label{eq:H_loc}
    \hat{H}_{\mathrm{loc.}} = \sum_{X\subset L}\hat{H}_X \ . 
\end{equation}
    \item \textbf{Gap:} In the thermodynamic limit $N\rightarrow \infty$, the energy difference between the first excitation and the ground state is finite.     
\end{enumerate}

Here, we will be interested in the properties in equilibrium, i.e., ground states (GSs). One of the main results of the \textit{section} is that the entanglement content of such states remains scalable:

\begin{overview}{Area law of entanglement}
Let $\ket{GS}$ be a ground state of a gapped local Hamiltonian $\hat{H}_{\mathrm{loc.}}$. The entanglement entropy $S_X$ with respect to a neighborhood $X$ scales with the area of the boundary of $X$, $|\partial X |$.    
\end{overview}

\noindent Notice that this is not what one usually expects for random pure many-body states. These would scale much more dramatically, with the volume $|X|$, $S_X\sim (N\log d -1)/2$, i.e. $S_X$ is extensive \cite{Page_1993, Bianchi2022}. The entropy growth with the system's size was first analyzed from field theory approaches~\cite{Srednicki1993}, also in the context of black-hole physics~\cite{Bombelli1986}. Within quantum information, the 1D area law was rigorously established first for Gaussian models~\cite{Audenaert2002}, and subsequently, under more general assumptions by Hastings~\cite{Hastings_2007}. For such systems, $S_X\sim \mathrm{constant}$. Hastings bound was improved in Refs.~\cite{Arad_2012,Arad_2013}. Another worth mentioning theorem is that exponential-decaying correlations imply area law \cite{Brandao_2014}. 
The extension of the previous results to higher dimensions is an open problem and a matter of intensive ongoing research \cite{Anshu_2022}. Thus far, the problem is addressed for Gaussian models~\cite{Plenio2005} and quasi-free fermionic and bosonic lattices~\cite{Cramer_2006}. Unexpectedly, the entropy scaling in such systems depends on the statistics (bosonic or fermionic).

The conditions imposed by the area law are quite stringent and only fulfilled by a set of states restricted to a \textit{corner} of the Hilbert space. In 1D those can be efficiently approximated by matrix-product states \cite{PerezGarcia_2006}: \\

\begin{figure}[h]
\sidecaption
% Use the relevant command for your figure-insertion program
% to insert the figure file.
% For example, with the graphicx style use
\includegraphics[width = .5\linewidth]{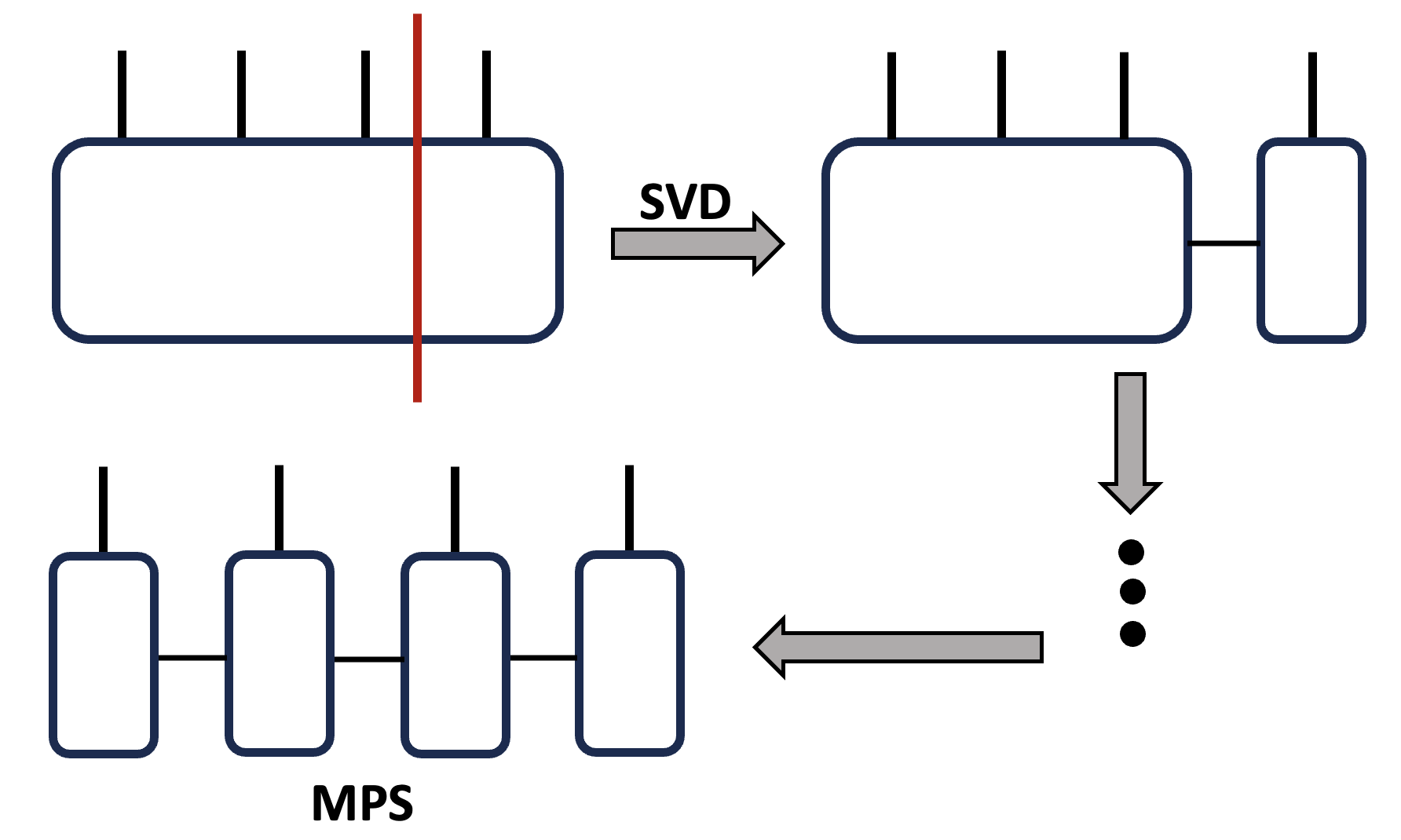}
%
% If no graphics program available, insert a blank space i.e. use
%\picplace{5cm}{2cm} % Give the correct figure height and width in cm
%
\caption{In the diagrammatic notation an $N$-tensor is represented as an $N$-legged box. Each leg represents a vector space of dimension $d$. By iterating SVD steps, one can factorize the tensor as an MPS.}
\label{fig:MPS}       % Give a unique label
\end{figure}

\noindent\textbf{Matrix product states}.-- Matrix product states (MPS) can be defined as a factorization in matrices (2-tensors) [c.f. Eq.~\eqref{eq:MB_sep}], 
\begin{equation}
\Psi_{s_1s_2..s_N}^{(D)}= \sum_{\{\alpha\in [D]\}} \psi_{s_1}^{\alpha_1}\psi_{s_2}^{\alpha_1\alpha_2}...\psi_{s_N}^{\alpha_{N-1}} = \vv{\psi}_{s_1}^T\psi_{s_2}...\vv{\psi}_{s_N} \ .
\end{equation}
Here, the scaling is $\order{D^2Nd}$. We refer to the size of such matrices, $D$, as the \textit{bond dimension}. For $D = 1$, we recover the separable case Eq.~\eqref{eq:MB_sep}. Note that any many-body state $\Psi$ can be factorized in this way by repeated singular value decompositions (SVD) for a sufficiently large bond dimension (see Figure~\ref{fig:MPS}). In the same Figure, we introduce the diagrammatic notation which is very useful to visualize tensor operations \cite{Vidal_2003,Vidal_2004}.  

From the picture above (Figure \ref{fig:MPS}), we learn that the bond dimension in a link is lower-bounded by the Schmidt rank with the respective bipartite cut. Consequently, for generic states, one has $D = d^{\lfloor N/2 \rfloor} \sim \mathrm{exp}(N)$. As the bipartite entanglement entropy is upper-bounded by the Schmidt rank:

\begin{svgraybox}
    Area law states in one dimension are well-approximated by matrix product states. 
\end{svgraybox}
\noindent Specifically, if we want to approximate the ground state within an error $\epsilon$ per spin, $||\ket{\mathrm{GS}}-\ket{\Psi^{(D)}}||\leq \epsilon/N$, a truncation of the bond dimension at $D\sim \mathrm{poly(N,\epsilon)}$ is sufficient \cite{Verstraete2006} (c.f. the exponential scaling for generic states). \\

MPS must be regarded as a variational ansatzes. There are more or less efficient algorithms to find an approximation of the $\ket{\mathrm{GS}}$ by optimizing the set of the many small tensors $\{\psi\}$. One of the most widely used is the density-matrix renormalization group (DMRG) \cite{White1992, Schollwck2005}. However, there are some Hamiltonians, whose ground state is exactly an MPS of constant bond dimension. Below, we give an example of it:

\begin{example}{Example: Affleck-Kennedy-Lieb-Tasaki (AKLT) model}
We place $N$ spin-1 in an open chain interacting according to the Hamiltonian \cite{Affleck1987}: 
\begin{equation}
\hat{H}_{\mathrm{AKLT}} = \sum_{i=1}^{N-1}\left[\hat{\mathbf{s}}_i\cdot\hat{\mathbf{s}}_{i+1} +\frac{1}{3} (\hat{\mathbf{s}}_i\cdot\hat{\mathbf{s}}_{i+1})^2\right] \ ,
\end{equation}
where $\mathbf{s}=(\hat{s}_x,\hat{s}_y, \hat{s}_z)$ is the vector of spin-1 matrices.

Here, we show how a bond dimension of $D=2$ is sufficient to represent the ground state of $\hat{H}_{\mathrm{AKLT}}$. In the following, we offer an explicit construction of it: \\

\noindent \textbf{Valence bond picture}.-- We embed the spin-1 local degrees of freedom into two auxiliary spin-$1/2$. Then, we deploy $N-1$ spin singlets $\ket{\psi^{-}}=(\ket{\uparrow\downarrow}-\ket{\downarrow\uparrow})/\sqrt{2}$ between bonds. The physical spin-1 is then recovered as it should be; after projection, $\mathcal{P}$, onto the symmetric sector of two adjacent spins. As sketched in Figure~\ref{fig:VBS}, such construction corresponds to an MPS of bond dimension $D = 2$: the-so-called valence bond state $\ket{\mathrm{VBS}}$ \cite{Fannes1991} and constitutes a GS of $\hat{H}_{\mathrm{AKLT}}$.

\begin{figure}[h]
\centering
% Use the relevant command for your figure-insertion program
% to insert the figure file.
% For example, with the graphicx style use
\includegraphics[width = .75\linewidth]{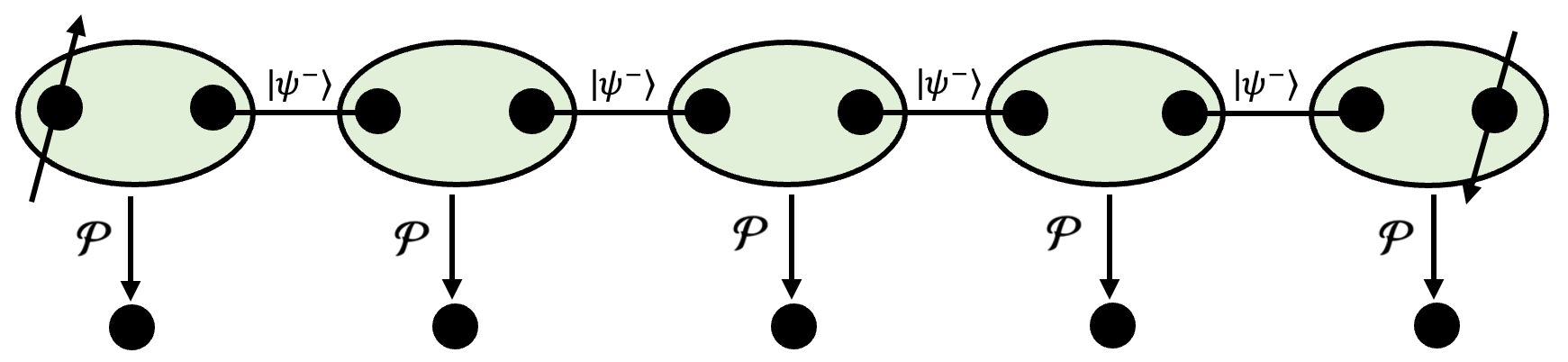}
%
% If no graphics program available, insert a blank space i.e. use
%\picplace{5cm}{2cm} % Give the correct figure height and width in cm
%
\caption{The GS is understood as an MPS by deploying spin singlets $\ket{\psi_-}$ between bonds and projecting, $\mathcal{P}$, adjacent sites to the physical space.  }
\label{fig:VBS}       % Give a unique label
\end{figure}

More generally, one can replace the singlet with any GME state in dimension $D$ and the symmetric projection with some linear map $\mathcal{P}: \mathbb{C}^D\otimes\mathbb{C}^D\rightarrow\mathbb{C}^d$. In fact, such construction is instrumental to implementing gauge invariance (e.g. for high-energy physics simulators \cite{Tagliacozzo2014}) and to generalizing MPS to higher dimensions through the so-called projected entangled pair states (PEPS) \cite{Verstraete2004a}.   \\

\noindent\textbf{Haldane phase and entanglement spectrum}.-- The VBS is a finite-size precursor of the Haldane phase, which, among others, is characterized by the presence of topological edge states (see Figure~\ref{fig:VBS}). In the thermodynamic limit, the entanglement spectrum [i.e., the eigenvalues of the (half chain) reduced density matrix] of the $\ket{\mathrm{VBS}}$ are exactly doubly-degenerate. As proven in Ref.~\cite{Pollmann_2010}, in spin-1 chains, such degeneracy can be lifted only if $(i)$ there is a quantum phase transition, $(ii)$ certain symmetries of the Hamiltonian are spontaneously broken. Consequently, such feature constitutes a robust signature of the Haldane phase, and more generally, it results in a tool to classify symmetry-protected topological phases in 1D, which was completed in Ref.~\cite{Chen2011}.      

\end{example}

\noindent\textbf{Gapless states and criticality}.-- Gap closing implies divergence of the correlation length\footnote{i.e., even though interactions are local, particles far apart become correlated.} in the system and may entail strong macroscopic consequences, like a critical point of a quantum phase transition. On the other hand, area-law-like theorems only provide sufficient conditions. The reverse is not true; e.g., there exist gapless Hamiltonians for which the GS entanglement content remains scalable. In particular, for some specific 1D systems, the bipartite entanglement entropy of a cut of length $L$ scales as $S_L\sim \log(L/a)$, where $a$ is the lattice spacing \cite{Calabrese2004}.\footnote{Note how it still scales exponentially better than a generic pure state.} Moreover, some efforts focused on the scaling of such quantity with the gap close to the critical point \cite{Gottesman_2010}. \\

\noindent\textbf{Quenches}.-- In this \textit{subsection} we have seen how MPS correctly reproduce the ground state properties of 1D local gapped Hamiltonians. One may ask whether MPS are useful for non-equilibrium states, such as quenches. This is generally not the case as entanglement increases rapidly with time, thus generating the volume law \cite{Calabrese_2005,Schuch2008}. However, the system may thermalize, and actually, quantum entanglement may no longer be necessary as correlations can be explained classically through mixing \cite{Leviatan_2017,Frias_2023}.     \\

To close the subject, we summarize in Table~\ref{tab:scaling} the different scaling laws outlined in this \textit{subsection}:    

\begin{table}[h]
\begin{center}
\begin{tabular}{|c| c |c|} 
 \hline
Family of states & Scaling of $S_X$ & Observations \\
 \hline
generic pure states & $|X|$ & curse of dimensionality   \\ 
GS of local, gapped Hamiltonians & $|\partial X|$ & proven for 1D system, approximable by MPS   \\ 
gapless states and criticality & $|\partial X| \log |X|$ & only for 1D specific cases  \\ 
quenches &   $|X|$ & for long times, however system may thermalize  \\
 \hline
\end{tabular}
\end{center}
\caption{Typical scaling of the bipartite entanglement entropy of region $X$ (versus the rest) for various classes of states of physical interest.}
\label{tab:scaling}
\end{table}

\subsubsection{Symmetric states and sufficient separability criteria}

Symmetries and invariance are of utmost importance in physics. They not only are instrumental in simplifying many problems but also fundamental to defining elementary interactions. Here we focus on symmetric states, i.e., those invariant under all permutation matrices.\footnote{Do not confuse \textit{symmetric} states by \textit{permutation-invariant} states. The latter refers to when the symmetry condition is relaxed to recover the same state up to a global phase, $\hat{T}\ket{\Psi} = e^{i\phi}\ket{\Psi}$, where $T$ is a permutation matrix.} The total symmetric sector of $N$ qudits, is spanned by the set \cite{Aloy2021}:

\begin{equation}
   \label{eq:def_sym}
    \left\{ \ket{S_\mathbf{k}} =\binom{N}{\mathbf{k}}
     \sum_{\pi \in \mathfrak{S}_{N}} \pi\left(\bigotimes_{a=0}^{d-1}\ket{a}^{\otimes k_a}\right)\right\}_{\mathbf{k}\vdash N}  ,
\end{equation}
where $\mathbf{k} = (k_0,k_1,..,k_{d-1})$ is a partition of $N$, ($\mathbf{k}\vdash N$), i.e. a set of positive integers adding to $N$ and the sum in the superposition is over unique permutations. The prefactor is the multinomial coefficient $\binom{N}{\mathbf{k}} = N!/\prod_{a=0}^{d-1}k_a!$.\footnote{However, the state is unnormalized, we may define the dual vectors $\ket{\tilde{S}_k} =\binom{N}{\mathbf{k}}^{-1} \ket{S_k} $ such that $\braket{\tilde{S}_k}{\tilde{S}_{k'}} = \delta_{kk'}$.}
The dimension of this subspace grows polynomially in $N$, with $\binom{N+d-1}{d-1}\sim\order{(N+d-1)^{d-1}/(d-1)!}$, which allows to process significant system sizes for low $d$.\footnote{For $d=2$, we will ease the notation and denote by $\ket{S_k}$, $\ket{S_{(k,N-k)}}$.} 

The symmetry constraints open the possibility for a scalable characterization of entanglement, leading to many unexpected results regarding the sufficiency of entanglement criteria. For instance \cite{Yu2016, Quesada2017},  
\begin{svgraybox}
Let $\hat{\rho}_{\mathrm{DS}}$ be a diagonal symmetric $N$-qubit state ($N$ even), i.e. of the form $\hat{\rho}_{\mathrm{DS}}=\sum_{k=0}^Np_k\ketbra{S_k}$, with $\{p_k\}$ a probability distribution. 

Then, $\hat{\rho}_{\mathrm{DS}}$ is separable if and only if $\hat{\rho}_{\mathrm{DS}}$ is PPT with respect to the largest bipartition $N/2:N/2$.
\end{svgraybox} 

\noindent Check Ref.~\cite{Rutkowski2019} for a generalization to qudits. 

Since the state is (in particular) permutation invariant, the PT does not depend on the specific parties considered but just on the total number of them. Moreover, the condition can be easily verified as $\hat{\rho}_{\mathrm{DS}}$ is PPT with respect to the bipartition $n:N-n$ iff the pair of Hankel matrices,

\begin{equation}
    \hat{M}_0(n) = \begin{pmatrix}
        p_0 & ... & p_n \\
        ... & ... & ...  \\
        p_n & ... & p_{2n}  \\
    \end{pmatrix} \  , \  \ 
        \hat{M}_1(n) = \begin{pmatrix}
        p_1 & ... & p_{n+1} \\
        ... & ... & ...  \\
        p_{n+1} & ... & p_{2n+1}  \\
    \end{pmatrix} \  , \  \ 
\end{equation}
are positive semidefinite (PSD).
The structure of such matrices also reveals that PPT with $n^*$ implies PPT of all bipartitions with $n\le n^*$ as they appear as principal minors. 

We will complete the exploration by providing a sketch of the proof of PPT sufficiency:  

\begin{proof} \cite{Yu2016}
By definition $\hat{\rho}_{\mathrm{DS}}$ is separable if for \textit{all} entanglement witnesses $\hat{W}$, $\mathrm{Tr}(\hat{W}\hat{\rho}_{\mathrm{DS}})\geq 0$. On the other hand, any entanglement witness detecting such states can be parametrized as a convex combination of:
\begin{align*}
    \hat{T} =   \sum_{0\leq i,j \leq N/2} a_ia_j \ketbra{\tilde{S}_{i+j}} \ , \  \ \hat{R} = \sum_{0\leq i,j \leq (N-1)/2} b_ib_j \ketbra{\tilde{S}_{i+j+1}} \ ,
\end{align*}
where $\mathbf{a}: = \{a_i \},\mathbf{b} := \{b_i\} $ are real coefficients. By applying them to our state, we obtain $\mathrm{Tr}(\hat{T}\hat{\rho}_{\mathrm{DS}}) = \mathbf{a}^T\hat{M}_0\mathbf{a}, \mathrm{Tr}(\hat{R}\hat{\rho}_{\mathrm{DS}}) = \mathbf{b}^T\hat{M}_1\mathbf{b}$.\footnote{Again, easing the notation $M_{0,1}:=M_{0,1}(N/2)$.} Such quantities are non-negative for all $\mathbf{a},\mathbf{b}$, if and only if $\hat{M}_{0,1}$ are PSD. Finally, as per the previous observation, PSD of the Hankel matrices  $\hat{M}_{0,1}$ implies PPT of $\hat{\rho}_{\mathrm{DS}}$ with respect to $N/2:N/2$.  
\end{proof}

\noindent\textbf{Permutation symmetry and spin}.-- For qubits, $(d=2)$ symmetric states are closely related to spatial rotations. Its associated group invariant is the total angular momentum (or spin), $\mathbf{S}^2 = S(S+1)$, where $S$ is the maximal spin projection. In such a setting, symmetric states have well-defined collective spin, $S=S_{\max} = Ns$, with $s=(d-1)/2 = 1/2$ the local spin, resulting from the coherent collective participation of the $N$ subsystems. The basis Eq.~\eqref{eq:def_sym} is then constructed via coarse-graining the manifold of total maximal spin by identifying vectors that can be related by a permutation among the $N$ subsystems (i.e., with the same spin projection). In higher dimensions $d>2$, symmetric states span different total spin sectors, from the macroscopic, giant spin $S = S_{\max}$ to the singlet $S = S_{\min} = 0$. This latter case corresponds to rotation invariant states.  \\

\noindent\textbf{Symmetric states in Nature}.-- The prototypical realization of symmetric states is the spinor Bose-Einstein condensate (e.g. of sodium or rubidium atoms) under the single-mode approximation \cite{Kawaguchi2012}. Similar collective models are realized under the denomination of Dicke models e.g. in cavity quantum-electrodynamics \cite{Dicke_1954} and in nuclear physics (as Lipkin shell models \cite{Likpin_1958}). These systems may display transitions between quantum chaotic and regular dynamics, which can also be explored from a Bell nonlocality perspective~\cite{Aloy2024chaos}. Symmetric selection rules are a standard way to scale quantum entanglement to the macroscopic regime. Hence, creating giant entangled states with depths of many thousands of atoms \cite{Lucke_2014, Zou2018}. As we show in the following, entanglement depth is a precious resource for sensing applications \cite{Pezze_2018}:  

\begin{overview}{Quantum-enhanced metrology}

Here, the goal is to infer the value of an unknown parameter $\theta$ (e.g., a magnetic field~\cite{Sewell_2012}, a space/time interval~\cite{Ligo_2019}, temperature~\cite{Stace_2010, Correa_2015, Srivastava_2023}). To this end, we employ a linear interferometer in order to encode the parameter as a phase in a quantum state via a unitary generated by $\hat{G}$, $\hat{\rho}(\theta) = e^{-i\theta \hat{G}}\hat{\rho}e^{i\theta \hat{G}}$ (see Figure~\ref{fig:Inter}). Finally, we measure an observable $O$.

\begin{figure}[h]
\centering
\sidecaption
% Use the relevant command for your figure-insertion program
% to insert the figure file.
% For example, with the graphicx style use
\includegraphics[width = 0.55\textwidth]{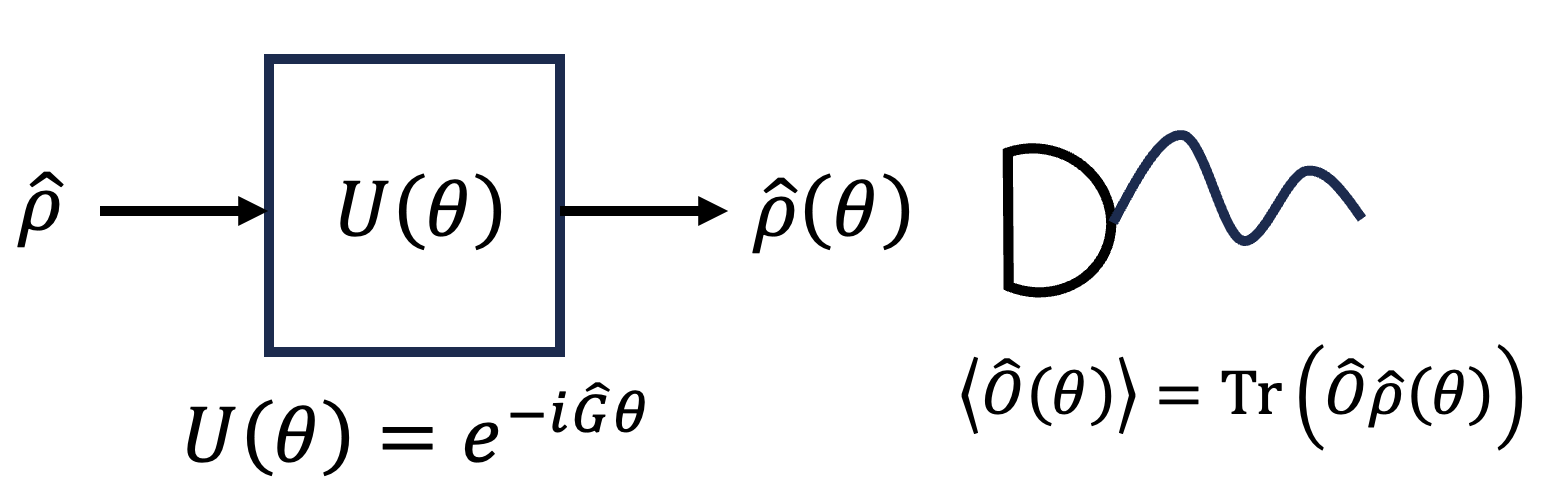}
%
% If no graphics program available, insert a blank space i.e. use
%\picplace{5cm}{2cm} % Give the correct figure height and width in cm
%
\caption{A parameter $\theta$ is encoded in a quantum state $\hat{\rho}(\theta)$ by linear interferometry, i.e. as a phase shift in a unitary manner: $\hat{\rho}(\theta) = e^{-i\theta \hat{G}}\hat{\rho}e^{i\theta \hat{G}}$. After the encoding, the state is measured against an observable $O$.  
}
\label{fig:Inter}       % Give a unique label
\end{figure}

After $\nu\rightarrow \infty$ number of shots, one can compute how the expectation value evolves in $\theta$, $\langle \hat{O}(\theta) \rangle$ and infer the value of $\theta$ from it. The precision of such a task is quantified by the variance of the estimator, $(\Delta\theta)^2_{\mathrm{est.}}$, which is bounded by the chain of inequalities \cite{Frowis2015,Gessner2019}:   

\begin{equation}
\label{eq:sens_bound}
\nu^2(\Delta\theta)^2_{\mathrm{est.}}\geq \frac{(\Delta \hat{O})^2_{\hat{\rho}}}{\langle i[\hat{G},\hat{O}] \rangle^2_{\hat{\rho}}}   \geq \min_{\hat{O}}\frac{(\Delta \hat{O})^2_{\hat{\rho}}}{\langle i[\hat{G},\hat{O}] \rangle^2_{\hat{\rho}}}:= \frac{1}{F_Q[\hat{\rho},\hat{G}]} \ .
\end{equation}
The ultimate bound $F_Q$ is the so-called quantum Fisher information (QFI) \cite{Braunstein1995}. As a central property, the QFI is a convex function of the state $\hat{\rho}$. 
States of high quantum Fisher information are used to enhance measurements mainly in quantum optics~\cite{Martos_2023}, for gravitational wave detectors \cite{Ligo_2019} or magnetometry with BECs \cite{Sewell_2012}.
Consider that we encode the phase collectively in $N$-qubits via a generator with structure $\hat{S}_z = \sum_{i\in[N]}\hat{\sigma}_{z,i}/2$, $\hat{\sigma}_z = \ketbra{\uparrow} - \ketbra{\downarrow}$.  
For such a case, using the convexity property, for any $K$-producible state $\hat{\rho}_K$ \cite{Hyllus2012,Toth2012}: 

\begin{equation}
    \label{eq:Fisher_depth}
F_Q[\hat{\rho}_K, \hat{S}_z]\leq NK \ .
\end{equation}

For $K=1$, the bound~\eqref{eq:Fisher_depth} is saturated by coherent spins states such as $[(\ket{\uparrow} + \ket{\downarrow})/\sqrt{2}]^{\otimes N}$ while for $K = N$ (a.k.a. the Heisenberg limit), the maximal sensitivity is achieved by the GHZ state $(\ket{\uparrow}^{\otimes N} + \ket{\downarrow}^{\otimes N} )/\sqrt{2}$.

The QFI is a highly nonlinear function of the states, which makes it challenging to infer experimentally. In Ref.~\cite{Palmieri_2023} machine-learning techniques ~\cite{dawid2022modern} were developed to infer such quantity in a realistic scenario. Likewise, Ref.~\cite{MullerRigat2023} puts forward theoretical tools to certify the metrological resource content of the state, as quantified by the QFI, if only partial information is available.   

\end{overview}

From inequality~\eqref{eq:Fisher_depth}, we learn:

\begin{svgraybox}
Increasing multipartite entanglement $K$ and/or the system size $N$ is necessary to improve the precision in phase estimation tasks.
\end{svgraybox}

\noindent Combining with Eq.~\eqref{eq:sens_bound} we build an \textit{entanglement depth witness}. Consider sensing collective rotations around the $z$-axis, $\hat{G} = \hat{S}_z$, and probing the spin projection in an orthogonal direction, $\hat{O}=\hat{S}_x$. Then, one is able to recover the Wineland spin squeezing criterion \cite{Wineland1994,Duan2002}:\footnote{We align the mean spin $\langle\hat{\mathbf{S}} \rangle$ along the $y$-axis such that $(\Delta\hat{S}_x)^2 = \langle \hat{S}_x^2\rangle$.} 
\begin{equation}
\label{eq:wineland}
\frac{\langle\hat{S}_x^2\rangle_{\hat{\rho}}}{\langle\hat{S}_y \rangle^2_{\hat{\rho}}}<\frac{1}{NK} \Longrightarrow\mbox{$\hat{\rho}$ has at least entanglement depth $K+1$},
\end{equation}
or stronger versions of it (check Refs.~\cite{Gessner2019,MullerRigat2023}). Precisely, the next section is devoted to the detection of quantum entanglement.  

\subsection{Scalable entanglement certification}

\begin{figure}[h]
\centering
%\sidecaption
% Use the relevant command for your figure-insertion program
% to insert the figure file.
% For example, with the graphicx style use
\includegraphics[width = 0.5\linewidth]{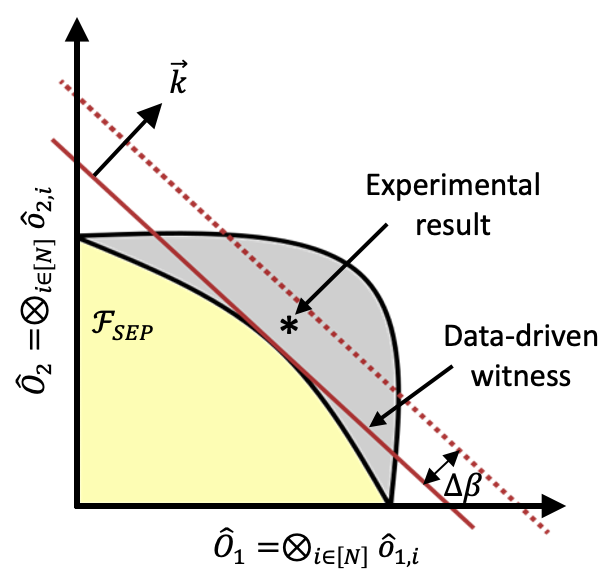}
%
% If no graphics program available, insert a blank space i.e. use
%\picplace{5cm}{2cm} % Give the correct figure height and width in cm
%
\caption{In yellow $\mathcal{F}_{\mathrm{SEP}}$, region of expectation values (correlations) compatible with a separable state. Such set is demarcated by (tight) entanglement witnesses (e.g. solid red line) characterized by the coefficients $\mathbf{k}$. The relaxation of the bound $\beta$ to $\beta' <\beta$ leads to a weaker witness (dashed red line), which may not be robust enough to detect some experimental data with genuine quantum origin (star). However, $\beta'$ can be less challenging to compute than $\beta$. Finally, the gray set contains all correlations reproducible by the quantum framework--but not classically recoverable.    }
\label{fig:witness_MB}       % Give a unique label
\end{figure}

\emph{As we have seen, quantum entanglement is omnipresent}. However, in the macroscopic scale, such feature is generally hidden in inaccessible higher-order correlations, and the world appears classical to us. With this, we highlight the importance of entanglement detection and the way we probe our systems. For instance, a state could be highly entangled; however, if we only measure commuting observables, the resulting correlations will always have a classical explanation. The same is true if only measure a correlator $\hat{O}_1 = \otimes_{i\in[N]}\hat{o}_{1,i}$. In such case, all quantum-feasible correlations $\langle\hat{O}_1\rangle$ are also reproducible classically -- with a separable state. However, if we combine two product observables $\hat{W} = k_1\hat{O}_1 + k_2\hat{O}_2$, as shown in Figure~\ref{fig:witness_MB} (c.f. Figure~\ref{fig:entanglement_witness}), then potentially some values of $\langle\hat{W}\rangle$ achievable with the quantum theory are not simulable classically. In the present subsection, we want to explore how to certify such fact in many-body systems, where already an in-depth reviews exists, Ref.~\cite{Frerot2023rev, Friis2018}.

To do so, we bound $\langle \hat{W} \rangle$ over separable states, i.e. find $\beta = \min_{\hat{\rho}\in \mathcal{R}_1}\mathrm{Tr}(\hat{W}\hat{\rho})$.\footnote{Note that by convexity of $\mathcal{R}_1$, optimization over pure states $\hat{\rho}=\ketbra{\Psi}$ is sufficient.} By convexity of the set of correlations accessible by separable states, $\langle \hat{W} \rangle_{\hat{\rho}}-\beta < 0$ implies $\hat{\rho}$ is entangled, i.e. entanglement is necessary to produce such statistics. \\

\noindent\textbf{The scalability issue}.-- Note that even in many-body systems, few correlators $\{\hat{O}_a\}_a:=\hat{\mathbf{O}}$ may be sufficient to detect entanglement, without the need for a full tomographic reconstruction of the state (which otherwise quickly becomes unfeasible with the system size). From a theoretical side, a challenge is to compute the bound $\beta$ (see Figure~\ref{fig:witness_MB}). In order to overcome such difficulty, one may relax the problem and instead aim to quantify weaker bounds $\beta'$, which are not tight but sufficient to detect entangled certain families of physically relevant many-body states.  \\

\noindent \textbf{The data-driven approach}.-- A further question that one may ask is the optimal witness (e.g. the best values of $\mathbf{k}=(k_1,k_2)$ of the previous discussion) to certify a given set of few experimentally-inferred\footnote{i.e. after partial (incomplete) tomography.} mean values $\langle \hat{\mathbf{O}}\rangle$, hereinafter called \textit{data}. Refs.~\cite{Frerot2021Bell,MullerRigat2021,Frerot_2022,MullerRigat2022, MullerRigat2023} provide generic tools to solve such problem in the many-body scenario for a particular class of witnesses. \\

Finally, from the technological aspect, entanglement is a pivotal resource, which is exploited in emergent quantum-enhanced applications. For this reason, the certification of such a feature is a necessary first step to assess the presumed advantage such devices may provide. The remainder of the subsection will be devoted to the detection of many-body quantum entanglement and Bell correlation.

\subsubsection{Detecting many-body entanglement in rotation-invariant states }

We start by giving an example of scalable entanglement witness by taking advantage of symmetries:  

\begin{example}{Example: Entanglement witness tailored to the many-body spin singlet}
Consider a system of $N$ spin-$s$ particles, from which we only can infer the expectation value of the total spin $\langle\hat{\mathbf{S}}^2 \rangle$. What can we say about the entanglement of the underlying quantum state? We can bound the expectation value over separable states Eq.~\eqref{eq:Kprod_mixed} $K=1$:
\begin{align}
\langle\hat{\mathbf{S}}^2 \rangle_{\hat{\rho}_{1}} &= \langle\left(\sum_{i\in[N]} \mathbf{s}_i\right)^2\rangle_{\hat{\rho}_{1}} = \underbrace{\sum_{i\in[N]}\langle\mathbf{s}_i^2\rangle_{\hat{\rho}_{1}}}_{=Ns(s+1)} + \underbrace{\sum_{i\neq j\in [N]}\langle\mathbf{s}_i\cdot \mathbf{s}_j\rangle_{\hat{\rho}_1}}_{\sum_{\psi}p_\psi \sum_{i\neq j\in [N]}\langle\mathbf{s} \rangle_{\psi_i}\cdot\langle\mathbf{s} \rangle_{\psi_j} } = \\
&= Ns(s+1) + \underbrace{\sum_\psi p_\psi\langle \sum_{i\in [N]}\mathbf{s}_i\rangle_{\psi} \cdot \langle \sum_{j\in [N]}\mathbf{s}_j\rangle_{\psi}}_{\geq 0}-\underbrace{\sum_{i\in [N]}\sum_\psi\langle\mathbf{s}_i\rangle^2_{\psi_i}}_{\leq Ns^2}\geq \\
&\geq Ns
\end{align}
Hence, for any $N$-partite separable state, 

\begin{equation}
    \label{eq:W_singlet}
    \langle\hat{\mathbf{S}}^2 \rangle\geq Ns \ .
\end{equation}
The witness Eq.~\eqref{eq:W_singlet} is maximally violated by the many-body spin singlet $\mathbf{S}^2 = 0$. Note that it does not detect metrologically useful entanglement, i.e., sensitivity on rotations, as there always exists a rotation invariant state compatible with $\langle\hat{\mathbf{S}}^2\rangle$ [c.f. Eq.~\eqref{eq:wineland}].    

\end{example}

Using the above example, one can ask about sufficiency. For instance, how can we characterize the set of mean spin $\langle\hat{\mathbf{S}} \rangle$ and its fluctuations (i.e., second moments) $\langle \hat{\mathbf{S}}\hat{\mathbf{S}}^T\rangle$ that are compatible with a separable state? Remarkably, for arbitrary $N$-partite spin-1/2 systems $(d=2)$, all entanglement that can be detected with such data can be summarized in only eight inequalities \cite{Vitagliano2014}. Such result was generalized to arbitrary spin \cite{Vitagliano2011, Vitagliano2014}. In Ref.~\cite{MullerRigat2022}, we formalize a data-driven approach to detect entanglement in spin ensembles with second moments of arbitrary collective observables (beyond spin projections). In such cases, we exploit Zeeman population measurements to unveil new witnesses tailored to relevant states prepared in spinor BEC experiments.

\subsubsection{Bell correlation}\label{subsec:Bell_correlation}

Bell correlations are one of the strongest tests of non-classicality. Here, in the so-called device-independent paradigm, we abandon any detailed assumption on the systems, e.g. if it can be described with quantum physics, the implementation of the measurements, etc. \cite{Scarani2013,Brunner2014}. The minimal description is encapsulated in only three numbers $(N,J,M)$. As before, $N$ is the number of parties in which we distribute a resource (e.g., a quantum state), $J$ is the number of measurements settings each party is able to choose, and $M$ is the number of measurement outcomes. After several rounds of collecting results, the parties can infer statistics, e.g. correlations in the form of joint expectation values.\footnote{Parties need to choose their settings for each round in a statistically independent way. This can be ensured by space-like separation. The lack of such a requirement makes LHV unfalsifiable. } We say that a correlation is classical if there exists a local hidden variable (LHV) model reproducing it. In other words, if it factorizes up to a \textit{hidden} statistical mixing:\footnote{Note that we deliberately removed the hats $\hat{\mathfrak{m}}$ from the measurements $\mathfrak{m}$ as quantum mechanics is not assumed. The expectation values should be understood in the purely statistical sense.  }  
\begin{equation}
    \label{eq:LHV}
    a,b\in [J],i,j\in[M]\ , \ \  \langle \mathfrak{m}_{a,i}\mathfrak{m}_{a,j}\rangle_{\mathrm{LHV}} = \sum_{\lambda} p_{\lambda}\langle \mathfrak{m}_{a,i}\rangle_{\lambda} \langle \mathfrak{m}_{b,j}\rangle_{\lambda}  \ ,
\end{equation}
where $\{p_{\lambda}\}$ is a probability distribution over the LHV $\lambda$. If for a given set of correlations (data), such probability does not exist, we denote them (Bell) nonlocal.\footnote{i.e., they require nonlocal hidden variables, beyond Eq.~\eqref{eq:LHV}.}   

It is surprising how such a simple description is sufficient to derive many nontrivial results. For instance, if quantum formalism is assumed, it is not difficult to show that all separable states cannot produce correlations beyond LHV. But the converse is not true, there are entangled states that do not violate any factorization of the form Eq.~\eqref{eq:LHV} \cite{Werner1989}. The set of LHV by construction forms a convex set. Consequently, nonlocality is signaled by a violation of a witness (a.k.a. Bell inequality) as in Figure~\eqref{fig:witness_MB}. From this discussion, we conclude:    

\begin{svgraybox}
Violating Bell inequalities allows one to certify the preparation of entangled states from minimal assumptions -- in a device-independent manner. 
\end{svgraybox}

Perhaps, more importantly, there exist (entangled) states which \textit{do} violate a Bell inequality \cite{Bell_1964}. With this fact, quantum formalism is justified as it derives predictions that cannot be classically reproduced. \\  

\noindent\textbf{Bell nonlocality versus Bell correlation}.-- In a standard Bell scenario, the inference of the necessary correlations to violate a multipartite Bell inequality quickly becomes intractable as the number of parties $N$ increases. Yet, for those that are permutation-invariant (PI), under some assumptions, it is not necessary to perform individual measurements as the very same correlations can be inferred via moments of collective observables.
For example, consider two parties and the quantum PI correlation $C_{ab} = 2(\langle \hat{a}\otimes \hat{b} \rangle + \langle \hat{b}\otimes \hat{a} \rangle)$. The inference of such expectation values seem to require individual addressing. However, it can also be expressed as $C_{ab} = \langle \hat{M}_a \hat{M}_b +\mathrm{h.c.}\rangle - \langle \hat{M}_{ab}\rangle$, where $\{
    \hat{M}_a = \hat{a}\otimes\mathbb{I} + \mathbb{I}\otimes\hat{a}, \
    \hat{M}_b = \hat{b}\otimes\mathbb{I} + \mathbb{I}\otimes\hat{b}, \ 
    \hat{M}_{ab} =  (\hat{a}\hat{b} + \hat{b}\hat{a})\otimes \mathbb{I} + \mathbb{I}\otimes (\hat{a}\hat{b} + \hat{b}\hat{a})\}$
are collective observables. Once the single-particle observables are specified, e.g. Pauli matrices $\{\hat{a}=\hat{\sigma}_x, \hat{b} = \hat{\sigma}_y \}$, then $\hat{M}_{ab}$ corresponds to a new observable (in this case $\hat{M}_{ab} = 0$). The two-particle term can be symmetrized $\langle \hat{M}_a \hat{M}_b +\mathrm{h.c.}\rangle = [\langle (\hat{M}_a + \hat{M}_b)^2\rangle - \langle (\hat{M}_a - \hat{M}_b)^2\rangle]/2   $. Thus, they may be evaluated by measuring $\hat{M}_a\pm\hat{M_b}$, corresponding to the collective spin in $x\pm y$ orientation.

The latter strategy is scalable in $N$. This is the natural measurement approach for quantum many-body systems, specifically cold atomic ensembles and Bose-Einstein condensates (BECs). However, in such systems, in addition, one can no longer guarantee space-like separation between parties.

Therefore, as we, (i) do not require closing loopholes, (ii) demand the validity of quantum mechanics, (iii) need a correct calibration of the settings e.g. spin orientation, we can no longer talk about Bell nonlocality. Instead, we use the term \textit{Bell correlations} to refer to the violation of multipartite Bell inequalities in a device-dependent way, from witnesses typically involving collective observables. From the assertions above, it is clear that Bell correlations are weaker than Bell nonlocality, but still they are instrumental to certify quantum entanglement. The detection of Bell correlations in many-body systems through collective measurements has been of great success in experiments with atomic ensembles \cite{Schmied2016,Engelsen2017} and in the context of quantum simulators \cite{Plodzien_2020,Plodzien_2022,Plodzien_2023, HernandezYanes_2022,HernandezYanes_2023,Dziurawiec_2023}. Beyond collective measurements, Bell correlation can also be detected from local data in 1D translation-invariant systems and other geometries \cite{Wang2017, Navascues2020, Navascues2021,Plodzien_2024_LMG}. 
Finally, it has been shown that as the number of particles grow, so does the violation of some Bell inequalities, like MABK~\cite{Mermin_1990,Ardehali_1992,Belinskii_1993} and WWW\.ZB~\cite{Werner_2001,Weinfurter_2001,Zukowski_2002} ones.\\

In order to derive the Bell inequalities, we need to bound functionals over LHV models, which might be challenging in the multipartite scenario. Below we outline a strategy to do so by interpreting the optimization as a statistical physics problem \cite{Frerot2021Bell} :    

\begin{overview}{Many-body Bell inequalities: a statistical physics approach}
Suppose we have a generic multipartite Bell inequality based on up to two-body correlation functions: 

\begin{equation}
    \mathcal{B} = \sum_{i\in[N]}\sum_{a\in [K]}k_{ai}\langle\mathfrak{m}_{a,i} \rangle + \sum_{i\neq j\in [N]}\sum_{a,b\in[K]}q_{ai,bj}\langle \mathfrak{m}_{a,i}\mathfrak{m}_{b,j}\rangle  \ ,
\end{equation}
where $\{k,q\}$ are real coefficients defining the Bell inequality. The goal is to find the classical bound, i.e. $\beta_{\mathrm{LHV}} = \min_{\mathrm{LHV}}\mathcal{B}$.     \\

\noindent\textbf{Local deterministic strategies}.-- From the convexity of the LHV set, for this task it is sufficient to consider extremal LHV models, i.e., deterministic strategies $\langle \mathfrak{m}_{a,i}\rangle = \sigma_{a,i}\in [M]$ (otherwise, any randomness is absorbed into the probability $\{p_{\lambda}\}$) and $\langle \mathfrak{m}_{a,i}\mathfrak{m}_{b,j}\rangle  =\langle \mathfrak{m}_{a,i}\rangle\langle \mathfrak{m}_{b,j}\rangle $~\cite{Pitowsky_1989}. Then, the problem can be reinterpreted as to find the ground energy of an Ising-like Hamiltonian, $\beta_{\mathrm{LHV}} = \min_{\{\sigma \in [M]\}} H$:

\begin{equation}
\label{eq:Ising_H}
    H = \sum_{i\in[N]}\sum_{a\in [K]}k_{ai}\sigma_{a,i} + \sum_{i\neq j\in [N]}\sum_{a,b\in[K]}q_{ai,bj}\sigma_{a,i}\sigma_{b,j} \ ,
\end{equation}
which can be solved with standard statistical-mechanical techniques such as simulated annealing, Monte Carlo, etc. 

Within the data-driven paradigm, the problem is not only to find the classical bound, but the value of the coefficients $\{k,q\}$ which (hopefully) detects the given data $\{\langle\mathfrak{m}_{a,i} \rangle , \langle \mathfrak{m}_{a,i}\mathfrak{m}_{b,j}\rangle \}$ Bell nonlocal. As shown in Ref.~\cite{Frerot2021Bell}, such a problem can be phrased as a convex optimization task with a well-defined solution.

As usual, in order to make the problem scalable, symmetries can be exploited. For instance, if one considers PI, that is $k_{ai} = k_a, q_{ai,bj} = k_{ab}$, the coarse-grained Hamiltonian Eq.~\eqref{eq:Ising_H} will only depend on the total number of parties which display outcome $r\in [M]$ when $a\in[J]$ is measured. 

\end{overview}

In the Bell scenario, the nature of the measurements remains unspecified. In a concrete physical situation, therefore, one has to optimize the settings to observe the maximal violation. This problem is, in general, highly non-convex and challenging in the multipartite regime. Even so, by exploiting symmetries, the first scalable Bell inequality derived in Ref.~\cite{Tura2014} led to the following Bell correlation witness \cite{Schmied2016}:

\begin{equation}
\label{eq:Bell_witness}
\frac{4\langle\hat{S}_x^2 \rangle}{N}< \frac{1}{2}\left[1-\sqrt{1-\left(\frac{2\langle \hat{S}_y\rangle}{N}\right)^2}\right]\Longrightarrow\mbox{Bell correlation} \ .
\end{equation}
Importantly, Ineq.~\eqref{eq:Bell_witness} is maximally violated by spin-squeezed states [c.f. Eq.~\eqref{eq:wineland}], but it is more demanding than the Wineland criterion Eq.~\eqref{eq:wineland}. In Ref.~\cite{MullerRigat2021}   Ineq.~\eqref{eq:Bell_witness} has been generalized to arbitrary local spin with a novel data-driven method which avoids the issue above. Similar witness are obtained for three-level multipartite Bell inequalities~\cite{Aloy2024, Muller2024}.   \\

We conclude the \textit{section} by highlighting that results like Eqs.~\eqref{eq:wineland}, \eqref{eq:W_singlet} and \eqref{eq:Bell_witness}, which are based on the same data, connect, on the same footing, three relevant quantum resources: metrology, entanglement, and Bell nonlocality~\cite{Niezgoda2021}. In particular, such data is inferred from experimentally accessible observables, e.g. those routinely exploited in ultracold atom platforms. Thus, eliciting the study of useful connections between quantum resources in realistic scenarios.

\section{Open problems}
\label{sec:5}
\textit{Here, we put forth a non-exhaustive list of open problems that were highlighted within the scope of this chapter. We hope that this discussion will rekindle the spark in the readers and the quantum information community to revisit these problems and address them in the coming years.}

\begin{enumerate}
    \item \textit{nPT bound entanglement}: All PPT entangled states are bound entangled~\cite{Horodecki_1998}. Is it also the other way around? In other words, are there bound entangled states that are not PPT? \\
    
    \item \textit{Classification of AME states}: Two AME states can be nonetheless LU-inequivalent \cite{Rather_2023}. The problem is to classify AME states in LU-inequivalent and/or other entanglement classes of interest. \\
    
    \item \textit{Sufficient conditions for multipartite LOCC convertibility}: As we have seen, Nielsen's theorem~\cite{Nielsen_1999} provides necessary and sufficient conditions for LOCC conversion in bipartite systems. In the multipartite setting, there are proposals to relax the conditions of LOCC to local separable (SEP) operations~\cite{Gour_2011}, by considering stabilizing symmetries~\cite{Sauerwein_2018}. While previous generalizations provide only necessary conditions, the quest for sufficient criteria of LOCC existence, beyond trivial symmetries, is open \cite{Nicky2023}.\\

    \item \textit{Unified theory of multipartite entanglement}: To this date, there does not exist a unified theory of multipartite entanglement despite the numerous efforts over the past three decades. This problem is intrinsically related to the mathematical problem of tensor classification~\cite{Bruzda_2023}. \\

    \item \textit{Higher dimensional area laws}: Rigorous proof of area law theorem/theorems in higher dimensions and suitability of PEPS in approximating higher dimensional ground states. Significant results are established for Gaussian models such as harmonic lattices   ~\cite{Plenio2005} and quasi-free fermionic and bosonic systems~\cite{Cramer_2006}. It would be beneficial to provide proofs under more general assumptions. \\
    
    \item \textit{Entanglement versus thermalization in many-body systems:} Under time evolution, entanglement generally increases rapidly. Such a scaling restricts the usefulness of many-body methods, like MPS, to short times. However, in typical scenarios, few-body correlations eventually thermalize and the necessary quantum entanglement to explain such statistics may remain nonetheless scalable. The task is to develop efficient standard tools to address thermalization via MPS or other low-entanglement ansatzes.  \\
   
    \item \textit{The separability problem in symmetric states}: As discussed in Section~\ref{sec:4}, there are no PPT entangled symmetric diagonal qubit states, which implies that the separability problem (in such subspace) is solved. Here, the task is to extend these results to qudit systems beyond the restricted set of Ref.~\cite{Rutkowski2019}, the bipartite case~\cite{Tura2018} or even diagonal states.  \\

    \item \textit{Metrological resources from a quantum information viewpoint}: Here, we highlight two questions:
    $(i)$ Does Bell nonlocality provide any advantage in the metrological task that is not explained by quantum entanglement, cf. Eq.~\eqref{eq:Fisher_depth}? 
    $(ii)$ As proved in Ref.~\cite{Toth_2018}, there exist PPT entangled states that nonetheless are metrologically useful. Which PnCP maps detect metrologically useful entanglement? 
    
\end{enumerate}

For a more detailed and elaborate list of open problems in quantum theory in general, please also look at Ref.~\cite{Horodecki_2022}, and the following links.\footnote{``\href{https://oqp.iqoqi.oeaw.ac.at/open-quantum-problems}{https://oqp.iqoqi.oeaw.ac.at/open-quantum-problems}'', ``\href{https://giedke.dipc.org/Benasque/futqi2023.pdf}{https://giedke.dipc.org/Benasque/futqi2023.pdf}''}

\newpage

\begin{acknowledgement}
We thank Karol \.{Z}yczkowski, Anna Sanpera, Ir\'en\'ee Fr\'erot, and Flavio Baccari for fruitful discussions and their valuable feedback.
% AKS:
A.K.S. acknowledges support from the European Union's Horizon 2020 Research and Innovation Programme under the Marie Sk{\l}odowska-Curie Grant Agreement No. 847517.
% ICFO:
ICFO group acknowledges support from:
ERC AdG NOQIA; MCIN/AEI (PGC2018-0910.13039/501100011033,  CEX2019-000910-S/10.13039/501100011033, Plan National FIDEUA PID2019-106901GB-I00, Plan National\newline STAMEENA PID2022-139099NB-I00 project funded by MCIN/AEI/10.13039/501100011033 and by the ``European Union NextGenerationEU/PRTR'' (PRTR-C17.I1), FPI); QUANTERA MAQS PCI2019-111828-2);  QUANTERA DYNAMITE PCI2022-132919 (QuantERA II Programme co-funded by European Union's Horizon 2020 program under Grant Agreement No 101017733), Ministry of Economic Affairs and Digital Transformation of the Spanish Government through the QUANTUM ENIA project call - Quantum Spain project, and by the European Union through the Recovery, Transformation, and Resilience Plan - NextGenerationEU within the framework of the Digital Spain 2026 Agenda; Fundaci\'{o} Cellex; Fundaci\'{o} Mir-Puig; Generalitat de Catalunya (European Social Fund FEDER and CERCA program, AGAUR Grant No. 2021 SGR 01452, QuantumCAT \ U16-011424, co-funded by ERDF Operational Program of Catalonia 2014-2020); Barcelona Supercomputing Center MareNostrum (FI-2023-1-0013); EU Quantum Flagship (PASQuanS2.1, 101113690); EU Horizon 2020 FET-OPEN OPTOlogic (Grant No 899794); EU Horizon Europe Program (Grant Agreement 101080086 - NeQST), ICFO Internal ``QuantumGaudi'' project; European Union's Horizon 2020 program under the Marie Sk{\l}odowska-Curie grant agreement No 847648;  ``La Caixa'' Junior Leaders fellowships, ``La Caixa'' Foundation (ID 100010434): CF/BQ/PR23/11980043. Views and opinions expressed are, however, those of the author(s) only and do not necessarily reflect those of the European Union, European Commission, European Climate, Infrastructure and Environment Executive Agency (CINEA), or any other granting authority.  Neither the European Union nor any granting authority can be held responsible for them. 
\end{acknowledgement}

\bibliography{references}

\end{document}